# Real Schur Flow Acoustic Directivity and Helicity Effects within

Zhu Min,[1] Pang Xu,[2] and Jian-Zhou Zhu (朱建州)[1, a)]

[1)] *Principle Department, Su-Cheng Center for Fundamental and Interdisciplinary Sciences, China*

[2)] *College of Meteorology and Oceanography, National Defence University, Changsha, 410073, China*

The statistical description and acoustic diagnostics of anisotropic turbulence remain challenging. Real Schur flows (RSFs), prototype models of component-wise dimensionally reduced flows (CWDRFs), provide effective platforms for studying such problems. We combine absolute statistical equilibrium with the acoustic analogy to investigate helicity effects on the far-field directivity of sound radiated from three types of CWDRFs: two different RSFs, and a further reduction of them (liberated Schur flow). We evaluate the directivity from the vortex-sound source spectrum which is calculated with the Wick theorem from the absolute equilibirum Gaussian ensemble. Helicity amplifies the dimensional reduction induced anisotropy of the directivity functions across all CWDRFs, with distinct patterns.

[a)] Electronic mail: jz@sccfis.org

## I. INTRODUCTION

The anisotropic-turbulence noise problem is much more challenging than that of the isotropic case, especially when it comes to the issue of particular effects of some physically important quantities such as helicity. Nevertheless, the study of acoustic directivity from turbulent flows has seen significant developments in recent years, driven by applications in aerospace engineering, urban air mobility, and environmental noise control. We briefly review the key advances closely relevant to the present work, organized by research theme.

*a. Helicity and vortex dynamics* Boutros and Gibbon[1] extended the definition of helicity to control non-conservation in compressible flows, offering new ways to characterize flow properties in high Reynolds number simulations. Liang et al.[2] applied vortex-sound theory to tornado infrasound emission, demonstrating that non-axisymmetric vorticity field rotation produces distinctive low-frequency acoustic signatures with characteristic directivity patterns.

*b. Swirling flow effects on acoustic modes* Faure-Beaulieu et al.[3] examined how swirling mean flows affect azimuthal aeroacoustic mode dynamics in annular combustors, demonstrating that controlled tangential injection creates helical velocity components that fundamentally alter modal stability and radiation patterns. Campos and Lau[4] investigated swirl Mach number effects on acoustic-vortical wave propagation, showing how swirling mean flows modify the coupling between acoustic and vortical modes.

*c. Cylinder and bluff body aeroacoustics* Huang et al.[5] studied heat exchanger tube bundle noise using a hybrid URANS-FWH method, finding that cross-flow direction noise levels exceed downstream direction values due to dipole source characteristics. Jiang et al.[6] examined square finite wall-mounted cylinders across a broad range of spanwise aspect ratios under favourable, zero, and adverse pressure gradients, identifying aspect-ratio-driven regimes that shift the Aeolian-tone spectrum and reshape the dipole-like far-field directivity pattern.

*d. Airfoil trailing-edge noise* Recent studies have advanced trailing-edge noise prediction. Ryu et al.[7] investigated turbulent boundary layer-trailing edge interaction noise for NACA0016 airfoils, analyzing far-field noise spectra and directivity patterns with distinct side-lobe behavior at high frequencies. Luo et al.[8] optimized trailing-edge shape for noise reduction using LES and ensemble-based methods, demonstrating how optimal shapes disrupt large-scale flow structures and suppress high-frequency noise.

*e. Computational aeroacoustic advances* Modern computational approaches have enabled detailed directivity predictions. Sharma et al.[9] combined LES with wind-tunnel experiments to characterize multi-scale turbulence structures and their directivity patterns. Prasad and Unnikrishnan[10] investigated plasma actuator-based shear layer control for rectangular jet noise mitigation, demonstrating that helical modes significantly influence acoustic radiation efficiency.

*f. Far-field directivity measurements* Advanced diagnostic methods have emerged for characterizing acoustic directivity fields. Kumar et al.[11] reported far-field noise measurements of a supersonic jet operating near afterburning conditions, mapping frequency-dependent dominant source regions along the jet axis and the angular redistribution of acoustic energy with operating condition. Botero-Bolívar et al.[12] studied trailing-edge noise under high inflow turbulence conditions, showing how wall-pressure fluctuations and spanwise correlation length variations lead to modified directivity patterns.

*g. Urban air mobility applications* Emerging applications have spurred new directivity research. Hanson et al.[13] experimentally investigated eVTOL rotor noise under reverse non-axial inflow conditions, finding that negative tilt angles significantly affect broadband noise distribution and directivity patterns, with implications for urban noise footprint prediction.

These recent studies collectively demonstrate that acoustic directivity from turbulent flows is intimately connected to coherent vortical structures, anisotropic mean flow effects, and the statistical properties of turbulence. However, a systematic theoretical framework, linking anisotropic structures to acoustic directivity and the helicity effect within, remains underdeveloped. These considerations constitute the main analysis of this paper.

In continuing the introductory discussion below, we first offer general remarks on component-wise dimensionally reduced flow (CWDRF) models that naturally offer the possibility of homogeneous anisotropic statistical mechanics without explicitly including (strong) constraints, acoustic analogy and statistical absolute equilibrium (Sec. I A). The theoretical foundations are subsequently presented: the Lighthill acoustic analogy framework (Sec. I A 1), and the mathematical formulation of component-wise dimensionally reduced flows (Sec. I A 2). We then articulate the specific research objectives and methodology (Sec. I B).

### A. CWDRF, acoustic analogy and statistical absolute equilibrium

The recently proposed real Schur flows (RSFs),[14,15] as prototype CWDRF models of anisotropic flows,[16] appear to provide a convenient and effective platform for studying various anisotropic turbulence problems, especially the challenging issues at the intersection of fluid mechanics and acoustics, such as statistical description and acoustic diagnostics.

The Lighthill acoustic analogy,[17–19] a classic and widely adopted approach in flow acous-tics, treats nonlinear problems using linear acoustics. Although the connection between such Lighthill-type 'sound' and what is 'heard' remains to be clarified further, to our point of view, the analogy nevertheless conveniently defines some 'acoustics' associated to the flow. Therefore, it is necessary to perform a Lighthill-type acoustic analogy analysis on the real Schur component-wise dimensionally reduced flows, both to check consistency and to obtain other insights through comparison.

For instance, in the RSF models of Refs. 14 and 15, the traditional mass equation was "truncated" (see below) in order to self-consistently maintain the velocity gradient matrix in the real Schur form in both space and time. Thus, one naturally asks: starting from such RSF equations, can one also derive the Lighthill acoustic analogy equation? If so, is it consistent with the classical acoustic analogy? This serves as an examination of the appropriateness of these RSF equations for anisotropic flow acoustics, and in turn, helps to reconsider the acoustic analogy itself.

The Lighthill acoustic analogy itself struggles to provide an explicit, unified theoretical treatment of the nonlinear problem as a whole. Other statistical approaches should be considered and, when possible, combined to be more effective in analysis. For example, T.-D. Lee (1952)[20] and R. Kraichnan (1955)[21] carried out related "absolute statistical equilibrium" studies for three-dimensional incompressible and compressible turbulence, respectively, obtaining non-trivial insights. Zhu[22] used helical decomposition to compute the statistical corrections due to helicity, updating the work of Ref. 21, and conjectured that helicity reduces flow compressibility (the "fastening effect"), and extended it to various plasma fluid models[23]. (The helicity fastening effect has been partially confirmed by spectral analysis of direct numerical simulations at low[15] and high[24] Reynolds numbers, encouraging us to continue theoretical exploration along this direction, especially in the RSFs: Ref. 15 actually already preliminarily showed the helicity fastening effect in a type of RSFs but investigations searching for a reasonable statistical mechanical support and further effects — such as the acoustic directivity — are still wanted.) Such analytical results cannot be obtained directly from the acoustic analogy. However, it is possible to "feed" the results of absolute statistical equilibrium calculations into the "source" part of the Lighthill equation to further examine consistency, thereby achieving a "theoretical closure".

### 1. *Acoustic analogy framework*

Let $\boldsymbol{u}$, $\rho$ and $c$ denote the velocity, density, and speed of sound of the flow, respectively, where the pressure $p$ is given by the following adiabatic barotropic relation,

$$p = c^2\rho, \quad \rho = \rho_0 e^{\zeta}. \tag{1}$$

By choosing appropriate scales and units, $c$ and the background density $\rho_0$ can be set to unity (= 1), but we will also let them appear explicitly when necessary to indicate physical meaning. The inviscid flow governing equations are

$$\partial_t \zeta + \zeta_{,\sigma} u_\sigma + u_{\sigma,\sigma} = 0, \tag{2}$$

$$\partial_t u_\lambda + u_\sigma u_{\lambda,\sigma} + c^2 \zeta_{,\lambda} = 0, \tag{3}$$

where $(\bullet)_{,\gamma} = \partial(\bullet)/\partial x^\gamma$. Consider a cubic periodic domain with volume $V = (2\pi)^3$. For any variable $v(\boldsymbol{r})$, its Fourier expansion coefficient is denoted $\hat{v}(\boldsymbol{k})$, with discrete wavevector $\boldsymbol{k}$. Sometimes, for convenience, we will also use a subscript notation $\hat{v}_{\boldsymbol{k}}$, and may even omit the wavevector when it is clear from context — for example, $\boldsymbol{u}(\boldsymbol{r}) = \sum_{\boldsymbol{k}} \hat{\boldsymbol{u}}(\boldsymbol{k}) \exp\{\hat{\imath}\boldsymbol{k} \cdot \boldsymbol{r}\}$, where $\hat{\imath}^2 = -1$ and the Fourier transform $\mathcal{F}[\boldsymbol{u}(\boldsymbol{x})](\boldsymbol{k}) = \hat{\boldsymbol{u}}(\boldsymbol{k})$.

Following Lighthill's derivation we obtain the exact acoustic analogy equation in terms of $\zeta$:

$$\frac{\partial^2 \zeta}{\partial t^2} - c^2 \nabla^2 \zeta = -\frac{\partial}{\partial t}(\boldsymbol{u} \cdot \nabla\zeta) + \nabla \cdot (\boldsymbol{u} \cdot \nabla \boldsymbol{u}). \tag{4}$$

The derivation of the acoustic-analogy equation can be carried out in more general settings, but the above setup simplifies the problem and allows us to focus on the core issues; we adopt the $\zeta$ variable to maintain consistency with the subsequent formulation of RSF equations and the absolute statistical equilibrium calculations.

### 2. *Real Schur component-wise dimensionally reduced flows*

We work in three-dimensional Euclidean space, so the relevant basic knowledge of matrix theory[26] is states that "all $3 \times 3$ real matrices can be turned into the 223- and 331-type real Schur ones by orthogonal (real Schur) transformations", with the following definitions: In a $3\times3$ matrix $R = \{r_{ij}\}$, if all other elements are arbitrary real numbers except that $r_{31} = r_{32} = 0$, then $R$ is called the 223-type real Schur matrix; if all other elements are arbitrary real numbers except that $r_{13} = r_{23} = 0$, then $R$ is called the 331-type real Schur matrix.

Now since $\nabla \boldsymbol{u}(\boldsymbol{x}, t)$ can always be locally (in space $\boldsymbol{x}$ and time $t$) transformed into the real Schur form, the corresponding CWDRFs — real Schur flows (RSFs) — based on such consideration of the "genericity" may be reminiscent of the local inertial frame and its role in connecting the special and general relativity theory and might be of fundamental importance. More specifically, CWDRFs are characterized by the uniformly vanishing of some component(s) of the velocity gradient matrix $\{u_{i,j}\}$ of the velocity $\boldsymbol{u} = \{u_1, u_2, u_3\} = \{\boldsymbol{u}_h, u_3\}$ ("$_h$" for the *horizontal plane* and "$_3$" for the *vertical direction*),

$$G = \begin{pmatrix} u_{1,1} & \boxed{u_{2,1}} & \cancel{u_{3,1}} \\ \enclose{circle}{u_{1,2}} & u_{2,2} & \cancel{u_{3,2}} \\ \xcancel{u_{1,3}} & \xcancel{u_{2,3}} & u_{3,3} \end{pmatrix}, \tag{5}$$

where an index behind the comma denotes the spatial derivative with respect to the corresponding coordinate variable. The corresponding flows, with the velocity gradient matrix uniformly (in space and time) of a particular real Schur form, are called real Schur flows (RSFs) or more generally component-wise dimensionally reduced flows (CWDRFs)[14–16].

Three types of CWDRFs are relevant here:

- **223RSF**: $\partial_{x_3} \boldsymbol{u}_h \equiv \boldsymbol{0}$ (as indicated by multiple slashes), i.e.,

$$u_{1,3} \equiv 0 \equiv u_{2,3}, \tag{6}$$

  so $u_1$ and $u_2$ are two-dimensional (2D2D), while $u_3$ remains three-dimensional (3D).

- **331RSF**: $\nabla_h u_3 \equiv \boldsymbol{0}$ (as indicated by single slash), i.e.,

$$u_{3,1} \equiv 0 \equiv u_{3,2}, \tag{7}$$

  so $u_3$ is one-dimensional (1D), varying only in $x_3$, while $u_1$ and $u_2$ remain three-dimensional (3D3D).

- **LSF (liberated Schur flow)**: an additional constraint $u_{2,1} \equiv 0$ is imposed on top of 331RSF, yielding a lower triangular velocity gradient matrix. Then $u_3$ is 1D (varies only in $x_1$), $u_2$ is 2D (in $x_3, x_2$), and $u_3$ is 1D — hence the designation 3D2D1D: switching the '1' and '3' indexes, we have the 1D2D3D LSF, thus the "uniqueness" of LSF. The LSF is "liberated" in the sense that it is free from closed streamlines (swirls); see below.

Associated to the barotropic Euler equation [Eqs. (2)–(3)], the self-consistent CWDRF equations, derived by requiring preservation of the respective real Schur form under the dynamics[16], are as follows. For **223RSF**:

$$\partial_t \zeta = \langle \varrho \rangle_{123} - \langle \varrho \rangle_3 - \langle \varrho \rangle_{12}, \tag{8a}$$

$$\partial_t \boldsymbol{u}_h + \boldsymbol{u}_h \cdot \nabla_h \boldsymbol{u}_h = -c^2 \nabla_h \zeta, \tag{8b}$$

$$\partial_t u_3 + \boldsymbol{u} \cdot \nabla u_3 = -c^2 \zeta_{,3}, \tag{8c}$$

where $\varrho := u_\sigma \zeta_{,\sigma} + u_{\lambda,\lambda}$, and $\langle \bullet \rangle_J$ denotes spatial averaging over the coordinate axes indexed by $J$. For **331RSF**, the $\zeta$ equation is identical [Eq. (8a)], but the momentum equations differ:

$$\partial_t \boldsymbol{u}_h + \boldsymbol{u} \cdot \nabla \boldsymbol{u}_h = -c^2 \nabla_h \zeta, \tag{9a}$$

$$\partial_t u_3 + u_3 u_{3,3} = -c^2 \zeta_{,3}. \tag{9b}$$

For **LSF**, the $\zeta$ equation and the momentum equations are further modified:

$$\partial_t \zeta = 2\langle \varrho \rangle_{123} - \langle \varrho \rangle_{23} - \langle \varrho \rangle_{13} - \langle \varrho \rangle_{12}, \tag{10a}$$

$$\partial_t u_1 + \boldsymbol{u} \cdot \nabla u_1 = -c^2 \zeta_{,1}, \tag{10b}$$

$$\partial_t u_2 + \boldsymbol{u}_h \cdot \nabla_h u_2 = -c^2 \zeta_{,2}, \tag{10c}$$

$$\partial_t u_3 + u_3 u_{3,3} = -c^2 \zeta_{,3}. \tag{10d}$$

*Key structural results.* We briefly summarize the main results from Ref. 16 that are relevant to the present acoustic analysis:

1. **No-go theorem (inequivalence of 223RSF and 331RSF).** There exists no uniform orthogonal transformation that turns all 331-type real Schur matrices into the 223-type (or vice versa). The two RSF types are therefore genuinely distinct: they have different momentum-equation structures [cf. Eqs. (8) and (9)] and different topological properties.

2. **Uniqueness of LSF.** By contrast, all upper-triangular $3 \times 3$ real matrices can be transformed into the lower-triangular form by a single uniform orthogonal transformation (e.g., the anti-diagonal permutation matrix swapping $x_1 \leftrightarrow x_3$). Hence there is only one type of LSF, and the 1D2D3D and 3D2D1D designations refer to the same flow class.

3. **Closed streamlines in 331RSF.** Since $u_3 = u_3(x_3, t)$ is a one-dimensional autonomous system along streamlines, $x_3$ is either monotonic or pinned at an equilibrium point $x_3^\#$ where $u_3(x_3^\#) = 0$. Consequently, closed streamlines in 331RSF can exist only on the horizontal equilibrium plane $x_3 = x_3^\#$.

4. **No swirls in LSF.** The cascading 1D2D3D structure of LSFs ($u_1 = u_1(x_1)$, $u_2 = u_2(x_1, x_2)$, $u_3 = u_3(x_1, x_2, x_3)$) forces each coordinate to be either monotonic or constant along any streamline, precluding any nontrivial closed streamline. LSFs are thus "vortical but with no swirls" — their vorticity is generally nonzero, yet no closed streamlines exist.

These structural differences among 223RSF, 331RSF, and LSF — in particular their distinct Fourier-mode supports and momentum-equation structures — are what give rise to the different directivity patterns analyzed below.

## B. Research objectives

Flow acoustic directivity and the helicity effects within naturally call for clarification of some basic notions, before laying out our research plan:

*a. Terminology clarification* In this paper, terms such as "anisotropy", "polarization", and "directivity" convey specific meanings that we clarify here (their compatibility with common terminology in flow acoustics will be further discussed later)[25]:

- "Anisotropy" refers generally to differences in different directions, as commonly understood.
- "Polarization" denotes that directional indicators (e.g., wavevector components rather than wavenumber) appear explicitly in a physical quantity, but not in a "dimensionally reduced" manner.
- "Directionality" specifically means that Kronecker delta functions representing dimensional reduction appear explicitly in the expression of physical functions, as shown later in Eqs. (32) and (34).

*b. Paper structure* In this paper, we extend the analysis beyond the 223RSF case to include three types of component-wise dimensionally reduced flows (CWDRFs): 223RSF, 331RSF, and 3D2D1D liberated Schur flow (LSF). These flows represent progressively more constrained anisotropic structures, allowing systematic investigation of how dimensional reduction affects helicity-induced acoustic directivity. In Sec. II, we derive the acoustic analogy equation for CWDRFs (Sec. II A), extend absolute statistical equilibrium analysis to these flows (Sec. II B 2), elucidate helicity effects (Sec. II B 3), and demonstrate the "directivity effect" across all three

CWDRFs (Sec. II C). Section III concludes with discussions (Sec. III A) and the outlook for future directions.

## II. ANALYSIS

The CWDRFs introduced in the introductory discussions are nothing but the mode-truncated versions of the original Euler equation, i.e., formally identical on the wavevectors support set, which leads to the satisfaction of accordingly the same properties, such as the Liouville theorem for the statistical mechanics analysis[27] and the derivation of the acoustic-analogy equation below: Since the truncation preserves the mode-interaction structure of the original system on the wavevector support set, such results are trivially obvious; but, it makes sense to formulate it explicitly, mainly for the specific case of 223RSF, without loss of generality, also for the purpose of getting familiar with the definite acoustic-analogy form for CWDRFs.

### A. Acoustic analogy for CWDRFs

For CWDRFs, the decomposability of $\zeta$ depends on the specific flow type. For 223RSF, $\zeta$ is non-zero only when $k_3 = 0$ or $\boldsymbol{k}_h = (k_1, k_2) = 0$. For 331RSF, $\zeta$ is non-zero only when $k_1 = 0$ or $k_2 = 0$ or $k_3 = 0$. For LSF, the support is further restricted. In all cases, the mass equation can be written as a natural truncation of the classical continuity equation, leading to formally identical acoustic analogy equations.

Taking the Fourier transform of the mass conservation equation and using the properties of averaging operators, we obtain for 223RSF:

$$\mathcal{F}[\langle \varrho \rangle_{123}](\boldsymbol{k}) \quad = \delta_{\boldsymbol{k},\boldsymbol{0}} \hat{\varrho}(\boldsymbol{0}), \tag{11}$$

$$\mathcal{F}[\langle \varrho \rangle_{12}](\boldsymbol{k}) = \delta_{\boldsymbol{k}_h,\boldsymbol{0}} \hat{\varrho}(0, 0, k_3), \tag{12}$$

$$\mathcal{F}[\langle \varrho \rangle_{3}](\boldsymbol{k}) \quad = \delta_{0,k_3} \hat{\varrho}(\boldsymbol{k}_h, 0), \tag{13}$$

which yields:

$$\frac{\partial}{\partial t}\hat{\zeta}(\boldsymbol{k}) = \delta_{\boldsymbol{k},\boldsymbol{0}} \hat{\varrho}(\boldsymbol{0}) - \delta_{\boldsymbol{k}_h,\boldsymbol{0}} \hat{\varrho}(0, 0, k_3) - \delta_{0,k_3} \hat{\varrho}(\boldsymbol{k}_h, 0). \tag{14}$$

For wavenumbers within the support of $\zeta$ (i.e., $k_3 k_h = 0$), only one term remains on the right-hand side of Eq. (14), and this term is exactly $-\hat{\varrho}(\boldsymbol{k})$ (even more explicit demonstration is given in the Appendix A). Therefore, on the support of $\zeta$, we have:

$$\frac{\partial}{\partial t}\hat{\zeta}(\boldsymbol{k}) = -\hat{\varrho}(\boldsymbol{k}), \text{ for } \boldsymbol{k} \text{ such that } k_3 k_h = 0 \tag{15}$$

This is precisely the Fourier form of the classical continuity equation. So, the RSF is nothing but the truncation of $\hat{\zeta}$ (and $\hat{\varrho}$) modes of $k_3 k_h \neq 0$ (a particular extremal of the Galekin truncations), which makes the subsequent derivation consistent with that of the classical untruncated flow.

### *1. Acoustic-analogy equation for CWDRFs in wavenumber space*

The derivation is simply that of classical hydrodynamics given in the introductory discussion, but now presented in the Fourier space.

Taking the Fourier transform of Eq. (3) yields:

$$\frac{\partial}{\partial t}\hat{\boldsymbol{u}}(\boldsymbol{k}) + \mathcal{F}[\boldsymbol{u}\cdot\nabla\boldsymbol{u}](\boldsymbol{k}) = -\hat{i}c^2\boldsymbol{k}\hat{\zeta}(\boldsymbol{k}). \tag{16}$$

Corresponding to the specific CWDRF constraints (Eqs. (8b, 8c) for 223RSF, and analogous equations for 331RSF and LSF), the velocity field constraints in Fourier space are expressed differently for each flow type. For 223RSF, we have:

$$\hat{\boldsymbol{u}}_h(\boldsymbol{k}) = \delta_{0,k_3}\hat{\boldsymbol{u}}_h(\boldsymbol{k}_h). \tag{17}$$

For 331RSF, the constraint is $\hat{u}_3(\boldsymbol{k}) = \delta_{0,k_1}\delta_{0,k_2}\hat{u}_3(k_3)$, and for LSF, additional constraints apply.

Taking the divergence (dotting with $\hat{i}\boldsymbol{k}$) of Eq. (16) gives:

$$\hat{i}\boldsymbol{k}\cdot\frac{\partial}{\partial t}\hat{\boldsymbol{u}} + \hat{i}\boldsymbol{k}\cdot\mathcal{F}[\boldsymbol{u}\cdot\nabla\boldsymbol{u}] = c^2|\boldsymbol{k}|^2\hat{\zeta}. \tag{18}$$

Taking the time derivative of Eq. (15) yields:

$$\frac{\partial^2}{\partial t^2}\hat{\zeta} = -\frac{\partial}{\partial t}\hat{\varrho}. \tag{19}$$

The Fourier expression of $\hat{\varrho}$ is given by $\varrho = \boldsymbol{u}\cdot\nabla\zeta + \nabla\cdot\boldsymbol{u}$:

$$\hat{\varrho}(\boldsymbol{k}) = \mathcal{F}[\boldsymbol{u}\cdot\nabla\zeta](\boldsymbol{k}) + \hat{i}\boldsymbol{k}\cdot\hat{\boldsymbol{u}}(\boldsymbol{k}). \tag{20}$$

Computing $\frac{\partial}{\partial t}\hat{\varrho}$ and using Eq. (18) to eliminate $\hat{i}\boldsymbol{k}\cdot\partial_t\hat{\boldsymbol{u}}$, we obtain:

$$\frac{\partial^2}{\partial t^2}\hat{\zeta} + c^2|\boldsymbol{k}|^2\hat{\zeta} = -\frac{\partial}{\partial t}\mathcal{F}[\boldsymbol{u}\cdot\nabla\zeta](\boldsymbol{k}) + \hat{i}\boldsymbol{k}\cdot\mathcal{F}[\boldsymbol{u}\cdot\nabla\boldsymbol{u}](\boldsymbol{k}) \text{ for } k_h k_3 = 0. \tag{21}$$

Noting that $\hat{i}\boldsymbol{k}\cdot\mathcal{F}[\boldsymbol{u}\cdot\nabla\boldsymbol{u}] = \mathcal{F}[\nabla\cdot(\boldsymbol{u}\cdot\nabla\boldsymbol{u})]$, we have:

$$\frac{\partial^2}{\partial t^2}\hat{\zeta}(\boldsymbol{k},t) + c^2|\boldsymbol{k}|^2\hat{\zeta}(\boldsymbol{k},t) = -\frac{\partial}{\partial t}\widehat{(\boldsymbol{u}\cdot\nabla\zeta)}(\boldsymbol{k},t) + \nabla\cdot\widehat{(\boldsymbol{u}\cdot\nabla\boldsymbol{u})}(\boldsymbol{k},t) \text{ for } k_h k_3 = 0. \tag{22}$$

This is the exact acoustic analogy equation for 223RSF in wavenumber space, structurally identical to the Fourier form of the $\zeta$ equation (4) for classical unconstrained flows. The difference lies in the fact that the field quantities $\boldsymbol{u}$, $\zeta$ are subject to the 223RSF constraints, resulting in a selective Fourier support.

### 2. *Vorticity form and modal decoupling*

Using the identity $\boldsymbol{u}\cdot\nabla\boldsymbol{u} = \nabla(|\boldsymbol{u}|^2/2) - \boldsymbol{u}\times\boldsymbol{\omega}$, where $\boldsymbol{\omega} = \nabla\times\boldsymbol{u}$, we have:

$$\nabla\cdot(\boldsymbol{u}\cdot\nabla\boldsymbol{u}) = \nabla^2\left(\frac{|\boldsymbol{u}|^2}{2}\right) - \nabla\cdot(\boldsymbol{u}\times\boldsymbol{\omega}) = \nabla^2\left(\frac{|\boldsymbol{u}|^2}{2}\right) + \nabla\cdot(\boldsymbol{\omega}\times\boldsymbol{u}). \tag{23}$$

Substituting into Eq. (22) gives:

$$\frac{\partial^2}{\partial t^2}\hat{\zeta} + c^2|\boldsymbol{k}|^2\hat{\zeta} = -\frac{\partial}{\partial t}\widehat{(\boldsymbol{u}\cdot\nabla\zeta)} + \widehat{\nabla^2\left(\frac{|\boldsymbol{u}|^2}{2}\right)} + \widehat{\nabla\cdot(\boldsymbol{\omega}\times\boldsymbol{u})} \text{ for } k_h k_3 = 0. \tag{24}$$

The constraints of 223RSF lead to the following characteristics of the Fourier support of the field quantities:

- When $k_3 = 0$ (horizontal plane modes), $\hat{\zeta} = \hat{\zeta}_h$, $\hat{\boldsymbol{u}}_h$ is non-zero, while $\hat{u}_3$ exists but is not restricted by $k_3 = 0$. Here, $\widehat{\nabla\cdot(\boldsymbol{\omega}\times\boldsymbol{u})}$ can be expanded in terms involving the vertical vorticity $\hat{\omega}_3$.

- When $\boldsymbol{k}_h = 0$ (vertical modes), $\hat{\zeta} = \hat{\zeta}_v$, $\hat{u}_3$ is non-zero, but $\hat{\boldsymbol{u}}_h = 0$, so $\hat{\omega}_3 = 0$.

- For other wavenumbers ($k_3 \neq 0, \boldsymbol{k}_h \neq 0$), $\hat{\zeta} \equiv 0$.

This modal dependence is the mathematical manifestation of the "directionality effect" in 223RSF.

### 3. *Powell–Howe approximation*

Comparing the exact CWDRF equation (22) with the $\zeta$ equation (4), we see that the forms are identical. The specificity of each CWDRF lies solely in the Fourier support of the field quantities, which gives rise to modal decoupling and helicity polarization effects.

Neglecting other terms and retaining only $\nabla\cdot(\boldsymbol{\omega}\times\boldsymbol{u})$ as the dominant sound source, and for small $\hat{\zeta} \approx \hat{p}/(\rho_0 c^2)$, multiplying by $\rho_0 c^2$ gives:

$$\frac{\partial^2\hat{p}}{\partial t^2} - c^2\nabla^2\hat{p} \approx \rho_0 c^2 \widehat{\nabla\cdot(\boldsymbol{\omega}\times\boldsymbol{u})} \text{ for } \boldsymbol{k} \text{ in the support of } \zeta, \tag{25}$$

which is the classical Powell–Howe equation[18,19]. By choosing appropriate units and normalization such that $c$ and $\rho_0$ take unit values, we can discuss $p$ and $\zeta$ interchangeably without formal distinction.

#### 4. *Clarification of terminology: anisotropy, polarization, directionality, directivity*

In line with the introduction, we reiterate and further clarify the following meanings:

- **Anisotropy**: A broad term referring to differences in physical quantities in different directions.

- **Polarization**: The explicit appearance of directional indicators (e.g., wavevector components $k_1, k_2$) in a physical quantity, but without involving dimensional reduction. In CWDRFs, statistics of certain modes depend explicitly on wavevector components, reflecting the polarization effect introduced by helicity. This resembles "directivity" in acoustics, but at the flow-field level.

- **Directionality**: Specifically, the explicit appearance of Kronecker delta functions (e.g., $\delta_{0,k_3}$) in expressions due to dimensional reduction. In 223RSF, $\hat{\boldsymbol{u}}_h$ exists only for $k_3 = 0$ modes; in 331RSF, $\hat{u}_3$ exists only for $k_1 = k_2 = 0$ modes; in LSF, additional constraints apply. These are manifestations of directionality.

- **Directivity**: Following the convention in aeroacoustics, this refers to the variation of observable acoustic quantities (such as pressure or particle velocity) with spatial direction. Our derivation shows that the polarization and directionality of the flow field ultimately map onto the directivity of the acoustic field.

This terminological framework helps to clearly describe the multi-level anisotropic features from flow statistics to acoustic observations.

## B. Absolute statistical equilibrium analysis of helicity effects in turbulence

#### 1. *A brief review*

Adopting a Galerkin truncation, e.g., setting to zero all modes with $k = |\boldsymbol{k}|$ greater than $K$, Kraichnan[21] considered the real and imaginary parts of $\hat{v}$ as constituting the phase space of the system, and it is easy to see that in the inviscid case the phase flow is incompressible, i.e., Liouville's theorem holds. For small perturbations, the mean energy per unit mass is

$$\mathcal{E} = \frac{\langle u^2 + c^2\zeta^2\rangle_{123}}{2} = \frac{\sum_{\boldsymbol{k}}[\hat{u}_\lambda(\boldsymbol{k})\hat{u}^*_\lambda(\boldsymbol{k}) + c^2|\hat{\zeta}(\boldsymbol{k})|^2]}{2}, \tag{26}$$

where the potential energy fluctuation is taken to second order

$$\int_{\rho_0}^{\rho} \frac{p - p_0}{\rho^2} d\rho \approx \frac{c^2 \zeta^2}{2}. \tag{27}$$

By considering the $H$-theorem, Kraichnan anticipated that the system would tend toward an absolute statistical equilibrium state. Starting from the canonical ensemble distribution $\sim \exp\{-\alpha\mathcal{E}\}$ (where $\alpha$ is a parameter related to "temperature"), he obtained equipartition of energy and used this to analyze the different dissipation rates that might lead to turbulence with or without noise.

Superficially, because the concept of statistical equilibrium involves infinite-time behavior, any approximation such as the potential energy above seems inadmissible as it would ultimately become uncontrolled. However, what we are actually interested in is the tendency of turbulent systems toward partial thermalization or thermalization over finite spatiotemporal scales[28]; therefore, within this finite physically relevant time, the reasonableness of the approximation cannot be ruled out. Moreover, although real turbulence, with dissipation and other factors, can never reach absolute statistical equilibrium, such qualitative trend analysis is not unreasonable over appropriate spatiotemporal scales. Due to the perservation of mode-interaction structures on the wavevector supports of the CWDRFs, as mentioned before, we will adopt the same approximation of energy conservation form of Kraichnan.[21]

At that time, it had not been known that this ideal system also conserves helicity[29],

$$\mathcal{H} = \sum_{\boldsymbol{k}} \hat{\imath}\boldsymbol{k} \times \hat{\boldsymbol{u}}_{\boldsymbol{k}} \cdot \hat{\boldsymbol{u}}_{\boldsymbol{k}}^{*}/2, \tag{28}$$

which was then used by Kraichnan in the absolute statistical equilibrium analysis of incompressible isotropic turbulence.[30] The issue of helicity in compressible flows has since been updated and supplemented[22] (more in Sec. II B 2), and the helicity conservation in CWDRFs[16] with further Galerkin truncation follows with the same argument.

#### 2. *Revisiting the case for isotropic turbulence*

Following Ref. 21, we consider a canonical ensemble distribution $\sim \exp\{-C\}$, but now with the constant of motion taken as $C = \alpha\mathcal{E} + \beta\mathcal{H}$, introducing a multiplier $\beta$, with $\mathcal{E}$ and $\mathcal{H}$ given directly by Eqs. (26) and (28), without employing the helical decomposition used in Ref. 22. Further separating the Fourier coefficients into real and imaginary parts, e.g., $\hat{u}_i(\boldsymbol{k}) = R_i + \hat{\imath}I_i$ $(i = 1, 2, 3)$, we obtain the correlation matrix for the tuple of variables $(R_1, R_2, R_3, I_1, I_2, I_3)$ (the part corresponding to $\zeta$ does not participate in any cross-terms in the quadratic form constituting $C$, so it only has an autocorrelation and is not listed):

$$\begin{pmatrix}
\frac{\alpha^2-\beta^2k_1^2}{\alpha^3-\alpha\beta^2k^2} & \frac{-\beta^2k_1k_2}{\alpha^3-\alpha\beta^2k^2} & \frac{-\beta^2k_1k_3}{\alpha^3-\alpha\beta^2k^2} & 0 & \frac{-\beta k_3}{\alpha^2-\beta^2k^2} & \frac{\beta k_2}{\alpha^2-\beta^2k^2} \\
\frac{-\beta^2k_1k_2}{\alpha^3-\alpha\beta^2k^2} & \frac{\alpha^2-\beta^2k_2^2}{\alpha^3-\alpha\beta^2k^2} & \frac{-\beta^2k_2k_3}{\alpha^3-\alpha\beta^2k^2} & \frac{\beta k_3}{\alpha^2-\beta^2k^2} & 0 & \frac{-\beta k_1}{\alpha^2-\beta^2k^2} \\
\frac{-\beta^2k_1k_3}{\alpha^3-\alpha\beta^2k^2} & \frac{-\beta^2k_2k_3}{\alpha^3-\alpha\beta^2k^2} & \frac{\alpha^2-\beta^2k_3^2}{\alpha^3-\alpha\beta^2k^2} & -\frac{\beta k_2}{\alpha^2-\beta^2k^2} & \frac{\beta k_1}{\alpha^2-\beta^2k^2} & 0 \\
0 & \frac{\beta k_3}{\alpha^2-\beta^2k^2} & -\frac{\beta k_2}{\alpha^2-\beta^2k^2} & \frac{\alpha^2-\beta^2k_1^2}{\alpha^3-\alpha\beta^2k^2} & \frac{-\beta^2k_1k_2}{\alpha^3-\alpha\beta^2k^2} & \frac{-\beta^2k_1k_3}{\alpha^3-\alpha\beta^2k^2} \\
-\frac{\beta k_3}{\alpha^2-\beta^2k^2} & 0 & \frac{\beta k_1}{\alpha^2-\beta^2k^2} & \frac{-\beta^2k_1k_2}{\alpha^3-\alpha\beta^2k^2} & \frac{\alpha^2-\beta^2k_2^2}{\alpha^3-\alpha\beta^2k^2} & \frac{-\beta^2k_2k_3}{\alpha^3-\alpha\beta^2k^2} \\
\frac{\beta k_2}{\alpha^2-\beta^2k^2} & -\frac{\beta k_1}{\alpha^2-\beta^2k^2} & 0 & -\frac{\beta^2k_1k_3}{\alpha^3-\alpha\beta^2k^2} & \frac{-\beta^2k_2k_3}{\alpha^3-\alpha\beta^2k^2} & \frac{\alpha^2-\beta^2k_3^2}{\alpha^3-\alpha\beta^2k^2}
\end{pmatrix}, \tag{29}$$

where $k = |\boldsymbol{k}|$. The six degrees of freedom from the real and imaginary parts of $\hat{\boldsymbol{u}}$ encompass

$$\text{the compressive mode } \hat{u}^{\parallel}\boldsymbol{k} \text{ and the vortical modes } \hat{\boldsymbol{u}}^{\perp} = \hat{u}^{+}\hat{\boldsymbol{h}}^{+} + \hat{u}^{-}\hat{\boldsymbol{h}}^{-}, \tag{30}$$

In the helical representation of Ref. 22, the former has two degrees of freedom (real and imaginary parts) in the direction of $\boldsymbol{k}$, while the latter consists of two pairs of real and imaginary degrees of freedom associated with the left and right helical eigenvectors ($\hat{\boldsymbol{h}}^{\pm}$): $\hat{u}^{\parallel}$ and $\hat{\zeta}$ have identical absolute statistical equilibrium properties (see Sec. II C below), but their imprints in real turbulence may differ significantly due to other factors (such as dissipation in Newtonian fluids) — this has been discussed in depth in previous works[21,22] and will not be repeated here.

The velocity spectral tensor is

$$\Phi_{ij}(\boldsymbol{k}) = \langle \hat{u}_i\hat{u}_j^*\rangle = \langle (R_i + \hat{\imath}I_i)(R_j - \hat{\imath}I_j)\rangle = \langle R_iR_j\rangle + \langle I_iI_j\rangle + \hat{\imath}(\langle I_iR_j\rangle - \langle R_iI_j\rangle), \tag{31}$$

where $\langle\bullet\rangle$ denotes statistical (ensemble) average. If we are only concerned with the difference in the total energy spectrum of fluctuations with helicity ($\beta \neq 0$) and without helicity ($\beta = 0$), a simple summation over components yields results consistent with Ref. 22. Moreover, the correlation function between $\hat{\zeta}$ and $\hat{\boldsymbol{u}}$ is zero (they are independent), and its autocorrelation takes the same form $1/\alpha$ whether helicity is present or not; its relative magnitude indeed decreases in the case with helicity (the "helicity fastening flow"), but this does not reflect the polarization effect we are concerned with here, so it is not listed above. Below we observe and discuss the polarization issue:

First, the most striking observation is that helicity introduces rich cross-correlation information beyond $\langle R_iI_i\rangle$: the fact that $\langle R_iI_i\rangle$ does not reflect helicity effects at all is because the right-hand side of Eq. (28) is actually equal to $\boldsymbol{k}\cdot(\boldsymbol{R}\times\boldsymbol{I})$. We see that helicity also induces polarization (on a spherical shell of constant $k$, the correlation functions become distorted as a function of direction). Such information is so diverse in the correlation functions of real and imaginary parts that its expected imprints in dissipative turbulence will also be rich. If we further decompose as in Ref. 22, we find that the compressive mode, like the density mode $\hat{\zeta}$, does not carry polarization effects either. That is, the latter is encoded only by the vortical modes.

It should be noted that not using the helical decomposition (30) is not only to facilitate comparison with the real Schur flow results below — which lack nontrivial pure helical modes[15] — but more substantially to reveal the helicity polarization effect: employing the helical decomposition (30) indeed brings computational convenience and cleanly yields the "fastening" effect; however, the helicity polarization effect is then completely hidden within the local helical coordinates $\hat{\boldsymbol{h}}^{\pm}_{\boldsymbol{k}}$ in wavevector space, so that polarization manifests in the difference between $\hat{u}^{+}$ and $\hat{u}^{-}$[22], rather than in the statistics of the coefficients $\hat{u}^{\pm}_{\boldsymbol{k}}$ themselves and their real and imaginary parts with respect to the direction of the wavevector. Thus, calculations in the two representations are complementary, and the current correlation matrix carries higher information content.

The polarization effect of helicity in isotropic turbulence is completely neutralized not only in "three-dimensional spectra" such as $\langle R^2(\boldsymbol{k}) = \sum_i R_i^2(\boldsymbol{k})\rangle$, but also, due to the $1:1:1$ spatial scales mentioned in the introduction, in one-dimensional cross-correlation spectra like $\langle\sum_{|\boldsymbol{k}|=k} R_1(\boldsymbol{k})R_2(\boldsymbol{k})\rangle$ (and consequently in physical-space statistics such as $\langle u_i u_j\rangle$), where it is also neutralized and not manifested: this is the significance of helicity polarization effects appearing in three-dimensional correlation spectra like $\langle R_i^2(\boldsymbol{k})\rangle$ and $\langle R_1(\boldsymbol{k})I_2(\boldsymbol{k})\rangle$ while the overall field remains isotropic. If the spatial scales are not $1:1:1$ (which is the general case in practice), then helicity polarization effects will also appear in cross-correlations, implying that in real flows, cross-correlations arising from other factors (if any) can be either amplified or suppressed by helicity.

### 3. *CWDRF absolute equilibrium analysis*

*a. 223RSF:* Equation (6) in Fourier space is expressed as $\hat{u}_1 k_3 \equiv 0 \equiv \hat{u}_2 k_3$, which implies

$$\hat{u}_1(1-\delta_{0,k_3}) \equiv 0 \equiv \hat{u}_2(1-\delta_{0,k_3}), \tag{32}$$

and leads to the correlation matrix for the tuple $(R_1, R_2, R_3, I_1, I_2, I_3)$ transforming from Eq. (29) to

$$\begin{pmatrix} \frac{\delta_{0,k_3}(\alpha^2-\beta^2k_1^2)}{\alpha^3-\alpha\beta^2k_h^2} & \frac{-\delta_{0,k_3}\beta^2k_1k_2}{\alpha^3-\alpha\beta^2k_h^2} & 0 & 0 & 0 & \frac{\delta_{0,k_3}\beta k_2}{\alpha^2-\beta^2k_h^2} \\ \frac{-\delta_{0,k_3}\beta^2k_1k_2}{\alpha^3-\alpha\beta^2k_h^2} & \frac{\delta_{0,k_3}(\alpha^2-\beta^2k_2^2)}{\alpha^3-\alpha\beta^2k_h^2} & 0 & 0 & 0 & \frac{\delta_{0,k_3}\beta k_1}{\beta^2k_h^2-\alpha^2} \\ 0 & 0 & \frac{\alpha}{\alpha^2-\delta_{0,k_3}\beta^2k_h^2} & \frac{\delta_{0,k_3}\beta k_2}{\beta^2k_h^2-\alpha^2} & \frac{\delta_{0,k_3}\beta k_1}{\alpha^2-\beta^2k_h^2} & 0 \\ 0 & 0 & \frac{\delta_{0,k_3}\beta k_2}{\beta^2k_h^2-\alpha^2} & \frac{\delta_{0,k_3}(\alpha^2-\beta^2k_1^2)}{\alpha^3-\alpha\beta^2k_h^2} & \frac{-\delta_{0,k_3}\beta^2k_1k_2}{\alpha^3-\alpha\beta^2k_h^2} & 0 \\ 0 & 0 & \frac{\delta_{0,k_3}\beta k_1}{\alpha^2-\beta^2k_h^2} & \frac{-\delta_{0,k_3}\beta^2k_1k_2}{\alpha^3-\alpha\beta^2k_h^2} & \frac{\delta_{0,k_3}(\alpha^2-\beta^2k_2^2)}{\alpha^3-\alpha\beta^2k_h^2} & 0 \\ \frac{\delta_{0,k_3}\beta k_2}{\alpha^2-\beta^2k_h^2} & \frac{\delta_{0,k_3}\beta k_1}{\beta^2k_h^2-\alpha^2} & 0 & 0 & 0 & \frac{\alpha}{\alpha^2-\delta_{0,k_3}\beta^2k_h^2} \end{pmatrix} \tag{33}$$

where $k_h^2 = k_1^2 + k_2^2$, and the Kronecker delta function $\delta_{0,k_3}$ reveals the directionality in the case with helicity ($\beta \neq 0$).

Like the other cases below, in the correlation function(s) vanishing with the Kronecker delta(s), the situation is actually that the corresponding (random) variable(s) being simply absent, in which case we must reduce the problem to avoid singularity; and, to formally unifythe presentation of results for both and inversions in a matrix, we have introduced Kronecker deltas to ensure that the (auto)correlations vanish for the corresponding variables absent.[27]

Mathematically, "polarization" manifests as the explicit inclusion of wavevector components such as $k_1, k_2$ in the matrix elements, implying that within the horizontal plane, the energy distribution varies continuously with direction; while "directionality" is embodied by the $\delta_{0,k_3}$ factor, implying a jump in statistical properties between the regions $k_3 = 0$ and $k_3 \neq 0$. Thus, "polarization is a continuous modulation with direction in wavevector space, while directionality is a discretized mutation due to dimensional constraints."

Beyond the discussions on helicity effects in isotropic turbulence similar to those in Sec. II B 2, a new feature worth noting here is that directionality and polarization often occur simultaneously; moreover, directionality also appears in the autocorrelations of physical quantities that are not explicitly dimensionally reduced, for example, the autocorrelations of $R_3$ and $I_3$:

$$\langle R_3^2 \rangle = \langle I_3^2 \rangle = \frac{\alpha}{\alpha^2 - \delta_{0,k_3}\beta^2 k_h^2}. \tag{34}$$

These are "equipartitioned" in all directions for $k_3 \neq 0$, but become more concentrated in the horizontal plane at $k_3 = 0$, and increasingly so with larger horizontal wavenumber $k_h$. This means that in this component-wise dimensionally reduced flow, although $u_3$ itself has no explicit dimensional reduction constraint, its fluctuations in the horizontal direction, especially at small scales, are enriched compared to the other two directions — in real turbulence, although small scales are subject to viscous dissipation, we have reason to expect this tendency to still occur. Here, directionality and polarization effects are intuitively difficult to distinguish, but according to our "definition" in the introduction, this anisotropy should be termed "directionality" (since $k_h = k$ when $\delta_{0,k_3} = 1$), although the practical significance of this distinction is limited.

The most obvious possible imprint in real turbulence of the correlation functions that are non-zero in the isotropic case of Eq. (29) but vanish here is that the corresponding quantities will be weakened due to the dimensional reduction feature.

*b. 331RSF:* Similarly, for the 331RSF, where $u_3$ is one-dimensional and varies only in $x_3$, the constraint in Fourier space is $\hat{u}_3 k_1 \equiv 0 \equiv \hat{u}_3 k_2$, which implies

$$\hat{u}_3(1 - \delta_{0,k_1}\delta_{0,k_2}) \equiv 0, \tag{35}$$

and leads to the correlation matrix

$$\begin{pmatrix} \frac{\alpha}{\alpha^2-\beta^2k_3^2} & 0 & 0 & 0 & \frac{-\beta k_3}{\alpha^2-\beta^2k_3^2} & 0 \\ 0 & \frac{\alpha}{\alpha^2-\beta^2k_3^2} & 0 & \frac{\beta k_3}{\alpha^2-\beta^2k_3^2} & 0 & 0 \\ 0 & 0 & \frac{\delta_{0,k_1}\delta_{0,k_2}}{\alpha} & 0 & 0 & 0 \\ 0 & \frac{\beta k_3}{\alpha^2-\beta^2k_3^2} & 0 & \frac{\alpha}{\alpha^2-\beta^2k_3^2} & 0 & 0 \\ \frac{-\beta k_3}{\alpha^2-\beta^2k_3^2} & 0 & 0 & 0 & \frac{\alpha}{\alpha^2-\beta^2k_3^2} & 0 \\ 0 & 0 & 0 & 0 & 0 & \frac{\delta_{0,k_1}\delta_{0,k_2}}{\alpha} \end{pmatrix}. \tag{36}$$

Compared to 223RSF, the directionality in 331RSF is manifested through $\delta_{0,k_1}\delta_{0,k_2}$ rather than $\delta_{0,k_3}$, reflecting the different component that is dimensionally reduced. The helicity polarization in 331RSF involves $k_3$ instead of $k_h$, yielding distinct anisotropic characteristics.

*c. LSF:* For the 3D2D1D LSF, which further imposes $u_{2,1} \equiv 0$ on top of 331RSF, the constraints are $\hat{u}_3(1 - \delta_{0,k_1}\delta_{0,k_2}) \equiv 0 \equiv \hat{u}_2(1 - \delta_{0,k_1})$, leading to the correlation matrix

$$\begin{pmatrix} \frac{\alpha}{\alpha^2-(\beta k_3)^2\delta_{0,k_1}} & 0 & 0 & 0 & \frac{-\beta k_3\delta_{0,k_1}}{\alpha^2-(\beta k_3)^2} & 0 \\ 0 & \frac{\alpha\delta_{0,k_1}}{\alpha^2-(\beta k_3)^2} & 0 & \frac{\beta k_3\delta_{0,k_1}}{\alpha^2-(\beta k_3)^2} & 0 & 0 \\ 0 & 0 & \frac{\delta_{0,k_1}\delta_{0,k_2}}{\alpha} & 0 & 0 & 0 \\ 0 & \frac{\beta k_3\delta_{0,k_1}}{\alpha^2-(\beta k_3)^2} & 0 & \frac{\alpha}{\alpha^2-(\beta k_3)^2\delta_{0,k_1}} & 0 & 0 \\ \frac{-\beta k_3\delta_{0,k_1}}{\alpha^2-(\beta k_3)^2} & 0 & 0 & 0 & \frac{\alpha\delta_{0,k_1}}{\alpha^2-(\beta k_3)^2} & 0 \\ 0 & 0 & 0 & 0 & 0 & \frac{\delta_{0,k_1}\delta_{0,k_2}}{\alpha} \end{pmatrix}. \tag{37}$$

The additional dimensional reduction in LSF introduces more $\delta_{0,k_1}$ factors, further suppressing helicity effects compared to 331RSF. Notably, even $\langle R_1^2\rangle$ and $\langle I_1^2\rangle$ contain the $\delta_{0,k_1}$ factor, despite $u_1$ having no explicit dimensional reduction — a consequence of the topological constraint that LSF has no closed streamlines (swirl).

### C. Helicity effects in absolute statistical equilibrium acoustic analogy

Directionality and polarization share a certain unity: for instance, in the aforementioned absolute statistical equilibrium, $\langle R_1^2\rangle$ (= $\langle I_1^2\rangle$) for the 223 real Schur flow can be understood as the polarization in isotropic flow with respect to $k_3$ becoming so strong that it is transformed into directionality characterized by $\delta_{0,k_3}$ in an extreme manner.

We can of course further compute other quantities of interest from the previous absolute statistical equilibrium results. For example, for the parallel modes ${}^{\parallel}\hat{u}_i = {}^{\parallel}R_i + \hat{\imath}\,{}^{\parallel}I_i := (\hat{\boldsymbol{u}}\cdot\boldsymbol{k})k_i/k^2$, from the isotropic case we obtain

$$\langle {}^{\parallel}R_i^2\rangle = \left\langle \left(\frac{\sum_{j=1}^3 R_j k_j}{k^2}\right)^2 k_i^2 \right\rangle = \frac{k_i^2}{\alpha k^2} = \langle {}^{\parallel}I_i^2\rangle, \tag{38}$$

while from the 223 real Schur flow we obtain

$$\langle {}^{\parallel}R_i^2\rangle = \left(\frac{\delta_{0,k_3}k_h^2}{\alpha} + \frac{k_3^2\alpha}{\alpha^2 - \delta_{0,k_3}\beta^2 k_h^2}\right)\frac{k_i^2}{k^4} = \langle {}^{\parallel}I_i^2\rangle. \tag{39}$$

From Eq. (38), we recover the polarization-independent result of Ref. 22:

$$\langle {}^{\parallel}R^2\rangle := \sum_{i=1}^{3}\langle {}^{\parallel}R_i^2\rangle = 1/\alpha. \tag{40}$$

Then,

$$\langle {}^{\parallel}R^2\rangle = \left(\frac{\delta_{0,k_3}k_h^2}{\alpha} + \frac{k_3^2\alpha}{\alpha^2 - \delta_{0,k_3}\beta^2 k_h^2}\right)\frac{1}{k^2} \tag{41}$$

encodes directionality and polarization through the explicit appearance of a Kronecker delta and of the wavevector components, respectively. We may also compute the energy of the vortical modes $\langle {}^{\perp}R^2\rangle := \left\langle \sum_{i=1}^3 R_i^2 - {}^{\parallel}R^2\right\rangle$:

$$\langle {}^{\perp}R^2\rangle = \frac{\delta_{0,k_3}\alpha}{\alpha^2 - \beta^2 k_h^2} + \frac{\alpha}{\alpha^2 - \delta_{0,k_3}\beta^2 k_h^2}\frac{k_h^2}{k^2}. \tag{42}$$

This expression exhibits a richer and more explicit mixture of directionality and polarization, including the concomitant "fastening" effect (relative enhancement of the vortical modes). Such intertwining greatly enriches the helicity-related information contained in component-wise dimensionally reduced flows.

The above comparison shows that the essential differences in the correlation matrices between the 223RSF and the isotropic case (e.g., the presence or absence of cross-terms such as $\langle R_1 R_2\rangle$) directly reflect different patterns of helicity polarization.

Similarly, for 331RSF, the parallel mode correlations are

$$\langle {}^{\parallel}R_i^2\rangle = \left(\frac{\alpha k_1^2}{\alpha^2-\beta^2 k_3^2} + \frac{\alpha k_2^2}{\alpha^2-\beta^2 k_3^2} + \frac{\delta_{0,k_1}\delta_{0,k_2} k_3^2}{\alpha}\right)\frac{k_i^2}{k^4} = \langle {}^{\parallel}I_i^2\rangle, \tag{43}$$

and the vortical mode energy is

$$\langle {}^{\perp}R^2\rangle = \frac{\alpha(1-k_1^2/k^2)}{\alpha^2-\beta^2 k_3^2} + \frac{\alpha(1-k_2^2/k^2)}{\alpha^2-\beta^2 k_3^2} + \frac{\delta_{0,k_1}\delta_{0,k_2}(1-k_3^2/k^2)}{\alpha} = \langle {}^{\perp}I^2\rangle. \tag{44}$$

For 3D2D1D LSF, the corresponding expressions are

$$\langle {}^{\parallel}R_i^2\rangle = \left(\frac{k_1^2}{\alpha} + \frac{\delta_{0,k_1}\alpha k_2^2}{\alpha^2-\beta^2 k_3^2} + \frac{\delta_{0,k_1}\delta_{0,k_2} k_3^2}{\alpha}\right)\frac{k_i^2}{k^4} = \langle {}^{\parallel}I_i^2\rangle, \tag{45}$$

and

$$\langle {}^{\perp}R^2\rangle = \frac{(1-k_1^2/k^2)}{\alpha} + \frac{\delta_{0,k_1}\alpha(1-k_2^2/k^2)}{\alpha^2-\beta^2 k_3^2} + \frac{\delta_{0,k_1}\delta_{0,k_2}(1-k_3^2/k^2)}{\alpha} = \langle {}^{\perp}I^2\rangle. \tag{46}$$

These results show that the progressive dimensional reduction from 223RSF to 331RSF to 3D2D1D LSF systematically reduces the helicity effects, with LSF exhibiting the weakest helicity polarization.

Such statistical anisotropy in wavenumber space, through the convolution inherent in the vortex-sound source term $\rho_0\nabla\cdot(\boldsymbol{\omega}\times\boldsymbol{u})$, will map onto differences in the "directivity" of the far-field acoustic radiation. Hence we now combine the acoustic analogy with the statistical results.

### *1. Source spectrum in vortex-sound theory*

For simplicity we restrict ourselves to the Powell–Howe approximation and consider the source term $s(\boldsymbol{x},t) = \nabla\cdot(\boldsymbol{\omega}\times\boldsymbol{u})$ and its spectrum $S(\boldsymbol{k}) := \langle|\hat{s}(\boldsymbol{k})|^2\rangle$:

$$\hat{s}(\boldsymbol{k},t) \;=\; \hat{i}\boldsymbol{k}\cdot\sum_{\boldsymbol{p}}\left[\hat{i}\boldsymbol{p}\times\hat{\boldsymbol{u}}(\boldsymbol{p},t)\right]\times\hat{\boldsymbol{u}}(\boldsymbol{k}-\boldsymbol{p},t), \tag{47}$$

$$\begin{aligned} S(\boldsymbol{k}) \;=\; \sum_{\boldsymbol{p}}\Big[&M^{\boldsymbol{k}}_{\alpha\beta}(\boldsymbol{p})M^{\boldsymbol{k}}_{\gamma\sigma}(\boldsymbol{p})\Phi_{\alpha\gamma}(\boldsymbol{p})\Phi_{\beta\sigma}(\boldsymbol{k}-\boldsymbol{p}) \\ &+M^{\boldsymbol{k}}_{\alpha\beta}(\boldsymbol{p})M^{\boldsymbol{k}}_{\gamma\sigma}(\boldsymbol{k}-\boldsymbol{p})\Phi_{\alpha\sigma}(\boldsymbol{p})\Phi_{\beta\gamma}(\boldsymbol{k}-\boldsymbol{p})\Big], \end{aligned} \tag{48}$$

where $M_{ij}^{\boldsymbol{k}}(\boldsymbol{m}) = k_i m_j - (\boldsymbol{k} \cdot \boldsymbol{m})\delta_{ij}$, which follows from the vector identity for $\nabla \cdot (\boldsymbol{\omega} \times \boldsymbol{u})$ and is formally valid for any velocity field. In Eq. (48) we have already used the Gaussian statistics of the absolute statistical equilibrium to express the fourth-order moments as the combination of the products of second-order moments (the Wick theorem: c.f., Appendix B for detailed calculations). The velocity spectral tensor $\Phi_{ij}(\boldsymbol{p})$ is given by Eq. (31); substituting it yields the source spectrum $S(\boldsymbol{k})$ in absolute statistical equilibrium, which can be used for demonstrating the helicity effects for given parameters $\alpha$ and $\beta$.

For component-wise dimensionally reduced flows, the same formal expression (48) applies, but with the velocity spectral tensor $\Phi_{ij}(\boldsymbol{p})$ replaced by the corresponding correlation matrices for each specific flow type (Eqs. (33), (36), and (37) for 223RSF, 331RSF, and 3D2D1D LSF, respectively). Crucially, however, the dimensional reduction constraints—manifested as Kronecker delta functions in the correlation matrices—radically alter the structure of the source spectrum beyond a simple substitution:

- Restricted summation domains. For 223RSF, since $\langle u_1(\boldsymbol{p})u_1^*(\boldsymbol{p})\rangle$ and $\langle u_2(\boldsymbol{p})u_2^*(\boldsymbol{p})\rangle$ are proportional to $\delta_{0,p_3}$, only wavevectors satisfying $p_3 = 0$ contribute to the $\boldsymbol{p}$-sum in Eq. (48) from the horizontal velocity components. Similarly, for 331RSF the constraint $\delta_{0,p_1}\delta_{0,p_2}$ on $\langle u_3 u_3\rangle$ restricts the vertical velocity contribution to the line $p_1 = p_2 = 0$, and for 3D2D1D LSF additional $\delta_{0,p_1}$ factors further restrict the summation.

- Vanishing cross-correlations. Different CWDRFs have different sets of non-vanishing cross-correlation functions (e.g., $\langle R_1 I_2\rangle$, $\langle R_2 I_1\rangle$). For the isotropic case, all six components are generally coupled; for 223RSF, the cross-terms involve $k_h$-dependent combinations; for 331RSF, only the $(R_1, I_3)$ and $(R_2, I_3)$ cross-correlations are non-zero; for 3D2D1D LSF, even fewer cross-terms survive due to the additional $\delta_{0,k_1}$ constraint. Consequently, the contraction of $M_{\alpha\beta}^{\boldsymbol{k}}(\boldsymbol{p})M_{\gamma\sigma}^{\boldsymbol{k}}(\boldsymbol{q})$ with $\Phi_{\alpha\gamma}(\boldsymbol{p})\Phi_{\beta\sigma}(\boldsymbol{q})$ yields different non-zero term structures for each flow type.

- Distinct polarization-directionality mixing. Because the helicity parameter $\beta$ couples to different wavevector components in each flow ($k_h$ for 223RSF, $k_3$ for 331RSF and LSF but with additional $\delta_{0,k_1}$ factors for the latter), the way helicity modulates the source spectrum is qualitatively different. In 223RSF, helicity-induced polarization is confined to the horizontal plane and modulated by $k_h$; in 331RSF, it involves the $k_3$ dependence of the horizontal components; in 3D2D1D LSF, the helicity effect is further suppressed by the $\delta_{0,k_1}$ constraint, reflecting the absence of swirl in LSF.

Thus, although the general Wick-theorem form (48) holds universally, the effective source spectrum—and hence the far-field directivity—depends nontrivially on the specific dimensional reduction structure through the interplay of the $M$-tensor geometry with the constrained velocity spectral tensor. The numerical results presented in Sec. III A are obtained by explicit evaluation of Eq. (48) with the corresponding correlation matrix for each flow.

#### 2. *Far-field approximation of the pressure autospectrum*

Consider statistically stationary turbulence. The space–time Fourier transform of the source term $s(\boldsymbol{x}, t) = \nabla \cdot (\boldsymbol{\omega} \times \boldsymbol{u})$ is defined as

$$\hat{\mathtt{s}}(\boldsymbol{k}, f) = \int s(\boldsymbol{x}, t) e^{-\hat{\imath}(\boldsymbol{k}\cdot\boldsymbol{x} - ft)} \, \mathrm{d}^3\boldsymbol{x} \, \mathrm{d}t.$$

Under the far-field condition $|\boldsymbol{x}| \gg$ source region extent, the frequency-domain pressure $\mathtt{p}(\boldsymbol{x}, f)$ (here and below, monospace letters denote frequency-domain quantities) can be approximated by the free-space Green function:

$$\mathtt{p}(\boldsymbol{x}, f) \approx \frac{e^{\hat{\imath}kr}}{4\pi r} \int \mathtt{s}(\boldsymbol{y}, f) e^{-\hat{\imath}k\tilde{\boldsymbol{n}}\cdot\boldsymbol{y}} \, \mathrm{d}^3\boldsymbol{y} = \frac{e^{\hat{\imath}kr}}{4\pi r} \hat{\mathtt{s}}(k\tilde{\boldsymbol{n}}, f), \tag{49}$$

where $k = f/c$ is the acoustic wavenumber, $r = |\boldsymbol{x}|$, and $\tilde{\boldsymbol{n}} = \boldsymbol{x}/r$ is the observation direction.

Defining the frequency autospectrum of the pressure as $\mathtt{S}_p(\boldsymbol{x}, f) = \langle |\mathtt{p}(\boldsymbol{x}, f)|^2 \rangle$ and substituting Eq. (49) yields

$$\mathtt{S}_p(\boldsymbol{x}, f) = \frac{1}{(4\pi r)^2} \langle |\hat{\mathtt{s}}(k\tilde{\boldsymbol{n}}, f)|^2 \rangle = \frac{1}{(4\pi r)^2} \mathtt{S}(k\tilde{\boldsymbol{n}}, f), \tag{50}$$

where $\mathtt{S}(\boldsymbol{k}, f) = \langle |\hat{\mathtt{s}}(\boldsymbol{k}, f)|^2 \rangle$ is the wavenumber–frequency spectrum of the source. Equation (50) shows that the far-field pressure autospectrum is proportional to the source spectrum evaluated at the acoustic wavevector $\boldsymbol{k}_s = k\tilde{\boldsymbol{n}}$, the proportionality factor depending only on the distance $r$. If we focus on the variation of the pressure with observation direction $\tilde{\boldsymbol{n}}$ at a fixed frequency $f$, we may simply write $\mathtt{S}(\boldsymbol{k}) \equiv \mathtt{S}(\boldsymbol{k}, f)$, where the wavenumber magnitude $k = |\boldsymbol{k}| = f/c$ is fixed by the frequency. Hence the directivity of the acoustic field is completely determined by the distribution of the spatial source spectrum $\mathtt{S}(\boldsymbol{k})$ on the sphere $|\boldsymbol{k}| = k$. *Actually, in the case of Gaussian fields delta-correlated in time, the frequency spectrum being independent of the frequency $f$, the wavenumber spectrum $S(\boldsymbol{k})$ is effectively the wavenumber-frequency spectrum $\mathtt{S}(\boldsymbol{k})$, which is particularly useful for understanding the far-field of acoustic-analogy results with our source spectrum computed from the absolute-equilibrium spectral tensor.* This relation directly links the flow statistics to the far-field acoustic observables and provides a theoretical basis for investigating the anisotropic effects on the directivity and further the helicity effects within.

### 3. *Directivity patterns and acoustic signatures of helicity effects*

The source spectrum $S(\boldsymbol{k}) = \langle |\hat{s}(\boldsymbol{k})|^2 \rangle$ given by Eq. (48) contains information about the spatial anisotropy of the acoustic source, including helicity-induced polarization and directionality. For a fixed wavenumber magnitude $k = |\boldsymbol{k}|$, the angular dependence of $S(k\tilde{\boldsymbol{n}})$ on the unit vector $\tilde{\boldsymbol{n}}$ directly defines the **wavenumber-domain directivity** of the source:

$$\mathcal{D}_{\text{wave}}(\tilde{\boldsymbol{n}}; k) = \frac{S(k\tilde{\boldsymbol{n}})}{\langle S(k\tilde{\boldsymbol{n}}) \rangle_{\tilde{\boldsymbol{n}}}}, \tag{51}$$

where $\langle \cdot \rangle_{\tilde{\boldsymbol{n}}}$ denotes averaging over all discrete wavevector directions consistent with the periodic box. This quantity is a pure measure of the source anisotropy, unaffected by propagation effects. According to the last remark in Sec. II C 2, Eq. (51) also makes sense for the far-field results in the frequency domain.

We have performed explicit numerical calculations on a discrete Fourier grid with cutoff $K = 51$, selecting wavevectors in the range $49.95 \leq k = |\boldsymbol{k}| \leq 50.05$ to approximate the shell $k \approx 50$, where the sphere is densely covered. The directivity $\mathcal{D}_{\text{wave}}(\tilde{\boldsymbol{n}}; k)$ was evaluated for multiple values of the dimensionless helicity parameter $\Gamma = \beta K/\alpha$ with $\alpha = 1$: with given $K$, $|\Gamma|$ increases with $|H|$ [the results do not depend on the signs of the helicity (parameters)]. The representative cases $\Gamma = 0$, $0.8$, and $0.98$ are collected in Fig. 1, arranged as a $3 \times 3$ array: the rows correspond to 223RSF [Figs. 1a–1c], 331RSF [Figs. 1d–1f], and 3D2D1D LSF [Figs. 1g–1i], and the columns correspond, from left to right, to the horizontal plane $k_3 = 0$ and the vertical planes $k_2 = 0$ and $k_1 = 0$. Each panel superposes the three helicity-parameter values using distinct markers (circles, stars, and open circles for $\Gamma = 0$, $0.8$, and $0.98$), and a dashed circle at radius 1 marks the isotropic baseline, which is accurately reproduced by the isotropic absolute equilibrium result for any helicity (and is therefore not shown as a separate curve). Consistency has been checked by other calculations using different $K$, $k$ and helicity parameters, thus the conclusions drawn below are of some generic sense.

*a. 223RSF* The computed patterns reveal that helicity enhances the anisotropy already present due to dimensional reduction, imprinting distinct signatures on the directivity. In the horizontal plane $k_3 = 0$ [Fig. 1a], the directivity already deviates [with the directivity function greater than 1, due to the fact that $u_h$ is varying only in the horizontal plane (with $k_3 = 0$)], "isotropically in this plane", from the isotropic baseline at zero helicity, and this deviation increases with helicity, as seen by the curves swelling nearly uniformly away from the dashed circle while remaining near-perfect circles (of radius $\approx 1.90$, 1.92, and 1.99 for $\Gamma = 0$, 0.8, and 0.98, respectively).

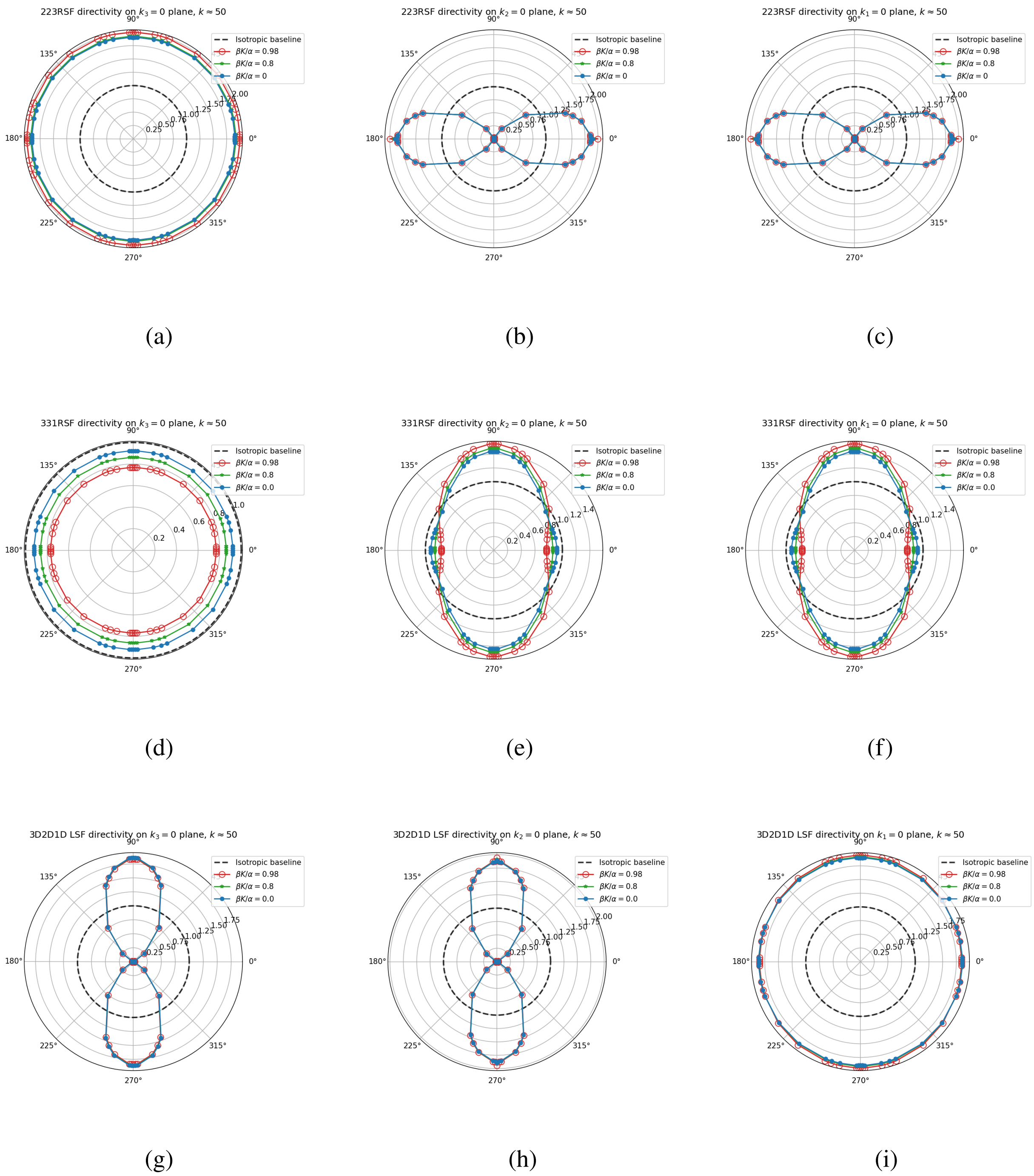

FIG. 1: Directivity patterns at $k \approx 50$ for $\beta K/\alpha = 0$ (circles), 0.8 (stars), and 0.98 (open circles) across all three CWDRFs. Top row: 223RSF; middle row: 331RSF; bottom row: 3D2D1D LSF. Each column shows a different plane: $k_3 = 0$ (left), $k_2 = 0$ (middle), and $k_1 = 0$ (right).

In the two vertical planes [Figs. 1b and 1c], the baseline (zero helicity) already exhibits a symmetric bimodal pattern along the $k_1$ and $k_2$ axes respectively, a direct consequence of the dimensional reduction. As helicity increases, the directivity at the two poles of $k_3 = 0$ becomes progressively more pronounced, while the rest of the pattern remains unchanged: only when $k_3 = 0$ the modes of $u_1$ and $u_2$ can enter to contribute. The equivalence of the $k_2 = 0$ and $k_1 = 0$ panels confirms the horizontal isotropy of 223RSF, as expected from the symmetric treatment of $u_1$ and $u_2$.

A technical detail in the numerical evaluation to clear the possible confusion follows. Note that the deviation of the curves from the dashed circle should be isotropic in the $k_3 = 0$ plane, i.e., presenting perfect circles of different radii — and indeed the computed curves are nearly perfect circles, the residual non-circularity ($\max/\min - 1$ of the directivity on this shell) being only 0.286%, 0.287%, and 0.418% for $\Gamma = 0$, 0.8, and 0.98, respectively — but in the numerical evaluation we have to take wavevectors from discrete lattices (here homogeneous integer grids), whose on-shell points are distributed non-uniformly in angle: for example, at the direction of $k_1 = 0$ or $k_2 = 0$ in this plane on the $k$-shell (here $49.95 \leq k \leq 50.05$, thus 52 wavevectors in it) we have smallest numbers (actually only one) of on-shell $\boldsymbol{k}$s which increase away from such extremal angles, so we have some minor "errors" depending on $k$, the angle, and the thickness of the $k$-shell etc.: we have tested such an issue by playing with different choices of them and presented the optimal results ("optimal" also with the consideration of the computational cost which obviously increases fastly with larger $k$ and $K$), but still the "errors" are unavoidable.

Many details depend on the choices of the parameters, but no essential points can be added concerning our purpose of this note. So, we won't perform a systematic survey of parameters here, but only showcase a particular one for better understanding: Fig. 2 presents the results at the shell with $39.92 \leq k \leq 40.08$ for $k \approx 40$ for 223RSF, with the cutoff $K = 51$ kept unchanged. We see that all the qualitative features established at $k \approx 50$ carry over intact to $k \approx 40$ — in the $k_3 = 0$ plane, near-perfect circles swelling with helicity, of radii $\approx 1.91$, 1.93, and 2.03 for $\Gamma = 0$, 0.8, and 0.98, respectively (a helicity enhancement of $\approx 6.5\%$, marginally stronger than the $\approx 4.9\%$ at $k \approx 50$); in the two vertical planes, the same bimodal patterns with the two poles of $k_3 = 0$ enhanced exactly as before and the $k_2 = 0$ and $k_1 = 0$ panels remaining indistinguishable; and deep, essentially helicity-insensitive minima at $(0, 0, \pm k)$ — up to detail-level quantitative shifts of precisely the discreteness type discussed above: the residual non-circularity, $\approx 0.92\%$, $0.92\%$, and $0.98\%$, is some two to three times that on the $k \approx 50$ shell, and the polar minima are filled in to $\approx 1.4$–$1.5 \times 10^{-2}$, as against $\approx 1.3$–$1.4 \times 10^{-4}$ at $k \approx 50$, the depth of such minima being a delicate cancellation and hence sensitive to the discreteness of the convolution sum. 331RSF and LSF are of similar fashion, thus not shown.

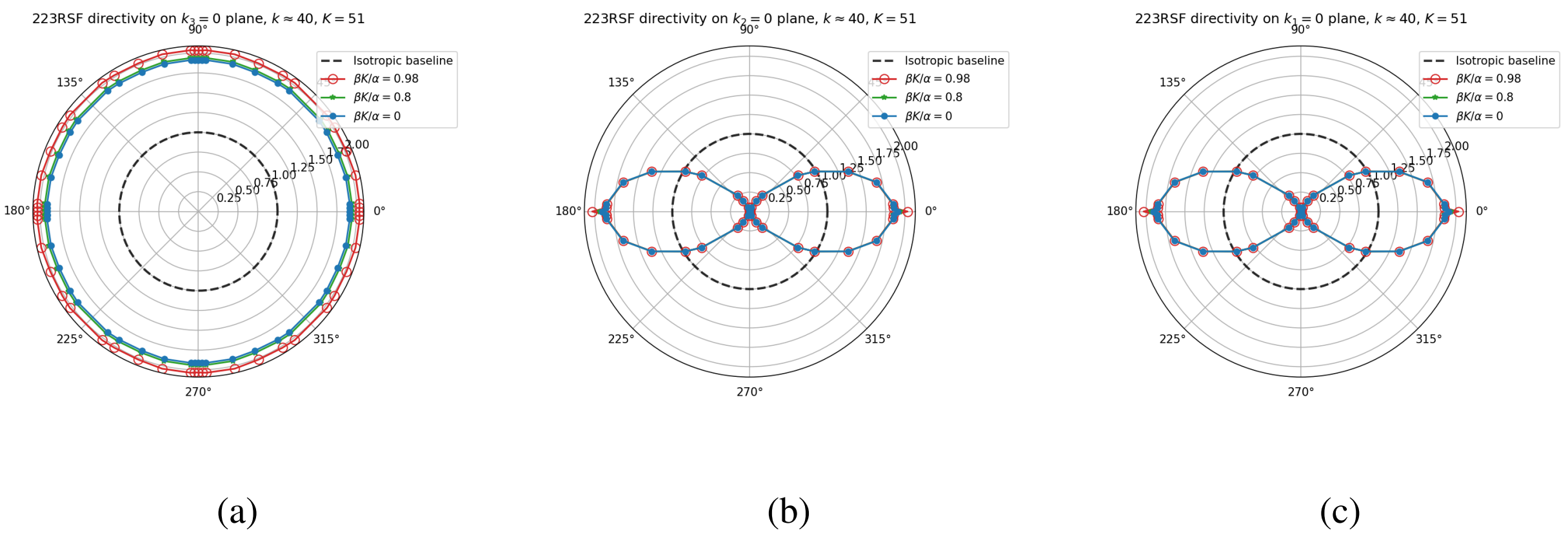

(a) (b) (c)

FIG. 2: 223RSF directivity at $k \approx 40$ (shell $39.92 \le |\boldsymbol{k}| \le 40.08$) for $\beta K/\alpha = 0$ (circles), 0.8 (stars), and 0.98 (open circles): $K = 51$. Columns as in Fig. 1: $k_3 = 0$ (left), $k_2 = 0$ (middle), and $k_1 = 0$ (right).

*b. 331RSF* For 331RSF, the directivity patterns exhibit distinct characteristics compared to 223RSF. In the horizontal plane $k_3 = 0$ [Fig. 1d], due to the fact that $u_3$ is varying only in the $x_3$ direction (with $k_3 \neq 0$), as apposed to the 223RSF case, the directivity function is smaller than 1, and more so with larger helicity. In the two vertical planes [Figs. 1e and 1f], the baseline already shows a different pattern compared to 223RSF, with the directional preference shifted toward the $k_3$ axis. As helicity increases, this preference becomes more pronounced, demonstrating that helicity amplifies the anisotropy inherent in the 331RSF structure. The equivalence of the $k_2 = 0$ and $k_1 = 0$ panels confirms the horizontal isotropy of 331RSF, as expected since both $u_1$ and $u_2$ are three-dimensional.

*c. 3D2D1D LSF* For 3D2D1D LSF, the directivity patterns show reduced helicity effects compared to both 223RSF and 331RSF, consistent with the additional dimensional reduction that further constrains the flow structure. In the planes $k_3 = 0$ [Fig. 1g] and $k_2 = 0$ [Fig. 1h] contain similar but quantitatively slightly different directivity functions, which can be similarly understood as in those 223 and 331RSFs. The most striking feature is the asymmetry between the $k_2 = 0$ and $k_1 = 0$ panels [Figs. 1h and 1i]: unlike the 223 and 331RSFs where these two planes are equivalent by horizontal isotropy, in LSF they differ qualitatively because the additional constraint $u_{2,1} \equiv 0$ breaks the symmetry between the $k_1$ and $k_2$ directions: for $k_3 = 0$, the $u_1$-difference between the $k_1$ ($x_1$) and $k_2$ ($x_2$) directions does not enter, thus the isotropicity in this plane. Most clearly, the $k_1 = 0$ panel [Fig. 1i] shows nearly constant isotropic directivity ($\approx 1.75$ with relative variation $< 5\%$), in stark contrast to 223RSF [Fig. 1a, $\approx 200\%$ of the baseline at $\Gamma = 0.98$] and 331RSF [Fig. 1d, $> 45\%$], providing a quantitative measure of how progressive dimensional reduction suppresses helicity effects. This reduced helicity effect and broken horizontal symmetry are direct consequences of the additional constraint $u_{2,1} \equiv 0$ in LSF, which further limits the flow's ability to respond to helicity.

Thus, the overall effect of helicity across all CWDRFs is to further accentuate the anisotropy originating from the dimensional reduction structure, with the magnitude of this enhancement depending on the specific type of CWDRF. 223RSF and 331RSF show stronger helicity effects compared to 3D2D1D LSF, reflecting the different degrees of dimensional reduction. The strong directional sensitivity of the far-field sound to the helicity parameter suggests that acoustic measurements, especially when combined with vector acoustic techniques, could serve as a non-invasive diagnostic tool for helicity in turbulent flows. The present work establishes a theoretical baseline for such diagnostics within the absolute statistical equilibrium framework.

## III. DISCUSSION AND OUTLOOK

### A. Discussion

Note that, for isotropic turbulence (which may still possess helicity, $\beta \neq 0$), the statistical isotropy implies that $\mathcal{D}_{\text{wave}}(\tilde{\boldsymbol{n}}; k) \equiv 1$ despite the presence of helicity. This does not mean that helicity has no effect on the acoustic field, but rather that its effect on the *scalar* pressure autospectrum is averaged out due to the isotropic averaging inherent in the definition of $S(\boldsymbol{k}) = \langle |\hat{s}(\boldsymbol{k})|^2 \rangle$. However, the correlation matrix (29) contains rich directional information even for isotropic turbulence: elements such as $\langle R_1 R_2 \rangle$, $\langle R_1 I_2 \rangle$, etc., exhibit explicit dependence on the wavevector components $k_1, k_2, k_3$. These finer statistical details are lost when only the scalar source spectrum $S(\boldsymbol{k})$ is considered.

To access this richer information experimentally, one must go beyond scalar pressure measurements and employ **vector acoustics**[31]. By synchronously measuring the pressure $p(\boldsymbol{x}, t)$ and the particle velocity vector $\boldsymbol{v}(\boldsymbol{x}, t)$, one obtains the full cross-spectral matrix of the acoustic

field:

$$\mathbf{\Phi}_v(\boldsymbol{x}, f) = \begin{pmatrix} \langle|\mathsf{p}|^2\rangle & \langle\mathsf{p}\mathsf{v}_1^*\rangle & \langle\mathsf{p}\mathsf{v}_2^*\rangle & \langle\mathsf{p}\mathsf{v}_3^*\rangle \\ \langle\mathsf{v}_1\mathsf{p}^*\rangle & \langle|\mathsf{v}_1|^2\rangle & \langle\mathsf{v}_1\mathsf{v}_2^*\rangle & \langle\mathsf{v}_1\mathsf{v}_3^*\rangle \\ \langle\mathsf{v}_2\mathsf{p}^*\rangle & \langle\mathsf{v}_2\mathsf{v}_1^*\rangle & \langle|\mathsf{v}_2|^2\rangle & \langle\mathsf{v}_2\mathsf{v}_3^*\rangle \\ \langle\mathsf{v}_3\mathsf{p}^*\rangle & \langle\mathsf{v}_3\mathsf{v}_1^*\rangle & \langle\mathsf{v}_3\mathsf{v}_2^*\rangle & \langle|\mathsf{v}_3|^2\rangle \end{pmatrix}. \tag{52}$$

In the far field, each component of this matrix can be related to the velocity spectral tensor $\Phi_{ij}(\boldsymbol{p})$ through the Green's function and the source expression. For example, the particle velocity components are obtained from the pressure gradient via the linearized momentum equation: $\boldsymbol{v} = (\hat{\imath} f \rho_0)^{-1}\nabla p$. In the frequency domain and far field, this yields

$$\mathsf{v}_i(\boldsymbol{x}, f) \approx \frac{\tilde{n}_i}{\rho_0 c}\mathsf{p}(\boldsymbol{x}, f) = \frac{\tilde{n}_i}{\rho_0 c}\frac{e^{\hat{\imath}kr}}{4\pi r}\hat{\mathsf{s}}(k\tilde{\boldsymbol{n}}, f). \tag{53}$$

Consequently, the cross-spectral density between two velocity components becomes

$$\langle\mathsf{v}_i(\boldsymbol{x}, f)\mathsf{v}_j^*(\boldsymbol{x}, f)\rangle = \frac{\tilde{n}_i\tilde{n}_j}{(\rho_0 c)^2(4\pi r)^2}\langle|\hat{\mathsf{s}}(k\tilde{\boldsymbol{n}}, f)|^2\rangle = \frac{\tilde{n}_i\tilde{n}_j}{(\rho_0 c)^2}\mathsf{S}_p(\tilde{\boldsymbol{n}}, f). \tag{54}$$

This shows that the velocity cross-spectrum is directly proportional to the pressure autospectrum, modulated by the directional factors $\tilde{n}_i\tilde{n}_j$. While this reproduces the same information as the scalar pressure, the full vector cross-spectral matrix contains additional independent components when considering cross-correlations between pressure and velocity:

$$\langle\mathsf{p}(\boldsymbol{x}, f)\mathsf{v}_i^*(\boldsymbol{x}, f)\rangle = \frac{\tilde{n}_i}{\rho_0 c}\mathsf{S}_p(\tilde{\boldsymbol{n}}, f). \tag{55}$$

More importantly, in the near field or when considering the full three-dimensional structure of the source, the vector cross-spectral matrix provides access to the phase information and directional correlations that are lost in the scalar pressure spectrum. For the purpose of this paper, we emphasize that the vector acoustic framework offers a pathway to directly measure the statistical anisotropies encoded in the correlation matrix (33), including those that are averaged out in the scalar source spectrum. By fitting the measured vector cross-spectral matrix to the predictions derived from the absolute statistical equilibrium model, one can in principle reconstruct not only the helicity parameter $\beta$ but also the detailed structure of the velocity correlations, including the polarization effects that are invisible in the scalar pressure directivity. This provides a powerful tool for experimental validation and for inverse problems in complex turbulent flows.

### B. Outlook

The present study opens several avenues for future research.

- **Beyond absolute equilibrium**: Real turbulent flows are never in absolute statistical equilibrium; they are affected by viscosity, finite Reynolds numbers, and non-equilibrium dynamics. Incorporating these effects, for instance through closures or direct numerical simulations, will allow us to assess the robustness of the equilibrium predictions and to identify which features survive in realistic conditions. The present work used the Gaussian approximation to close the fourth-order moments. Future work could explore corrections due to non-Gaussianity, which may become important at high Reynolds numbers or in the presence of strong intermittency.

- **Experimental validation**: The predicted directivity patterns, especially the extreme directional lobes at high helicity, could be tested in laboratory settings using well-controlled swirling flows or in atmospheric/oceanic acoustics where helicity may be present. Vector acoustic sensors offer a practical means to capture the full cross-spectral matrix, enabling direct comparison with the theoretical predictions.

- **Inverse problem**: The strong dependence of directivity on $\beta$ suggests a potential inverse procedure: by measuring the far-field acoustic directivity and fitting it to the equilibrium model, one could estimate the helicity level of the source region. This could be particularly valuable for monitoring turbulent flows where direct velocity measurements are difficult.

In conclusion, the framework established here seem to provide a nice theoretical foundation for understanding and exploiting anisotropic turbulence (polarization, directionality and acoustic directivity) and the helicity effects within. And, we anticipate that it will stimulate further developments in both fundamental turbulence theory and practical acoustic diagnostics. But, as a theoretical approach, several caveats should be pointed out: first, as always, the absolute equilibrium only indicates the possible trend(s) of the relevant physical properties in turbulence (with dissipation),[32] here the CWDRF turbulence, so it must be iterated that our results indicate only that the corresponding turbulence may tend to have such aeroacoustic directivities; second, anisotropy, specifically here the truncations of the modes of some velocity component(s) down to $k_i = 0$ for particular component(s) $k_i$ of the wavevector $\boldsymbol{k}$, can introduce additional time scale(s) that may deteriorate the relevance of the thermalization (whose time scale is supposed to be comparable to the eddy turn-over time of isotropic turbulence); third, the CWDRF relevance to realistic flow dynamics, such as the 331RSF to stratification, is not very clear yet;[16] and, fourth, the Lighthill-type acoustic analogy is but an analogy our calculation has so far been restricted to the source strength (power spectrum), not directly the object that is heard/measured.

The list continues, but it is enough to wind up by saying that we should go in this direction bravely, but carefully.

## ACKNOWLEDGMENTS

We thank the discussions with H. Xiang and S. Xiong.

## DATA AVAILABILITY

The data that support the findings of this study are available from the corresponding author upon reasonable request.

## Appendix A: Consistency with the classical form of the $\zeta$ equation

For wavevectors within the support of $\zeta$, two cases are considered:

- If $k_3 = 0$ and $\boldsymbol{k}_h \neq 0$, then $\hat{\zeta}(\boldsymbol{k}) = \hat{\zeta}_h(\boldsymbol{k}_h)$, and $\delta_{\boldsymbol{k},0} = 0$, $\delta_{\boldsymbol{k}_h,0} = 0$, $\delta_{0,k_3} = 1$; hence the right-hand side of Eq. (14) becomes $-\hat{\rho}(\boldsymbol{k}_h, 0)$. From Eq. (20), $\hat{\varrho}(\boldsymbol{k}_h, 0)$ is exactly $\mathcal{F}[\boldsymbol{u} \cdot \nabla\zeta + \nabla \cdot \boldsymbol{u}]$ evaluated at $(\boldsymbol{k}_h, 0)$. Therefore $\partial_t\hat{\zeta}_h = -\hat{\varrho}(\boldsymbol{k}_h, 0)$.
- If $\boldsymbol{k}_h = 0$ and $k_3 \neq 0$, then $\hat{\zeta}(\boldsymbol{k}) = \hat{\zeta}_v(k_3)$, and $\delta_{\boldsymbol{k},0} = 0$, $\delta_{\boldsymbol{k}_h,0} = 1$, $\delta_{0,k_3} = 0$; hence the right-hand side is $-\hat{\varrho}(0, 0, k_3)$, i.e., $\partial_t\hat{\zeta}_v = -\hat{\varrho}(0, 0, k_3)$.

Thus, on the support of $\zeta$, we always have $\partial_t\hat{\zeta} = -\hat{\varrho}$, which is consistent with the classical form.

## Appendix B: Derivation of the vortex-sound source spectrum [Eq. (48)]

We detail the derivation of $S(\boldsymbol{k}) = \langle|\hat{s}(\boldsymbol{k})|^2\rangle$ [Eq. (48)] from the Powell–Howe source $s(\boldsymbol{x}, t) = \nabla\cdot(\boldsymbol{\omega}\times\boldsymbol{u})$. Three steps are essential: (i) the convolution structure of $\hat{s}$ in wavenumber space, (ii) the cancellation of two Levi-Civita symbols that yields the bilinear $M$-tensor, and (iii) the combination of fourth-order moments via Wick's theorem together with the wavenumber selection rules that single out the two surviving terms.

### Step 1. Convolution form of $\hat{s}$ [Eq. (47)]

For homogeneous turbulence, differentiation becomes multiplication ($\widehat{\nabla f} = \hat{\imath}\boldsymbol{k}\,\hat{f}$) and a physical-space product becomes a wavenumber-space convolution ($\widehat{fg}(\boldsymbol{k}) = \sum_{\boldsymbol{p}} \hat{f}(\boldsymbol{p})\hat{g}(\boldsymbol{k}-\boldsymbol{p})$).

Since $\hat{\omega}_n(\boldsymbol{p}) = [\hat{\imath}\boldsymbol{p} \times \hat{\boldsymbol{u}}(\boldsymbol{p})]_n = \hat{\imath}\,\varepsilon_{nab}\, p_a \hat{u}_b(\boldsymbol{p})$, the divergence $\nabla \cdot (\boldsymbol{\omega} \times \boldsymbol{u})$ transforms as

$$\hat{s}(\boldsymbol{k}, t) = \hat{\imath}\, k_m\, \varepsilon_{mnl} \sum_{\boldsymbol{p}} \hat{\omega}_n(\boldsymbol{p}, t)\, \hat{u}_l(\boldsymbol{k}-\boldsymbol{p}, t) = \hat{\imath}\, \boldsymbol{k} \cdot \sum_{\boldsymbol{p}} \left[\hat{\imath}\boldsymbol{p} \times \hat{\boldsymbol{u}}(\boldsymbol{p}, t)\right] \times \hat{\boldsymbol{u}}(\boldsymbol{k}-\boldsymbol{p}, t),$$

which is Eq. (47). The first (index) form is the one used below; the second is recovered by recognizing the contraction $\hat{\imath}\, k_m \varepsilon_{mnl} A_n B_l = \boldsymbol{k} \cdot (\boldsymbol{A} \times \boldsymbol{B})$.

#### Step 2. Levi-Civita cancellation: emergence of the $M$-tensor

Substituting $\hat{\omega}_n = \hat{\imath}\varepsilon_{nab} p_a \hat{u}_b$ and collecting the prefactor $\hat{\imath}\cdot\hat{\imath} = -1$ gives

$$\hat{s}(\boldsymbol{k}) = -k_m\, \varepsilon_{mnl}\varepsilon_{nab} \sum_{\boldsymbol{p}} p_a \hat{u}_b(\boldsymbol{p})\hat{u}_l(\boldsymbol{k}-\boldsymbol{p}).$$

The **key cancellation** is the contraction of the two $\varepsilon$ symbols over the shared index $n$. Relabelling $\varepsilon_{mnl} = \varepsilon_{nlm}$ (a cyclic, hence even, permutation) places the summed index first in both factors, so that the identity $\varepsilon_{nab}\varepsilon_{ncd} = \delta_{ac}\delta_{bd} - \delta_{ad}\delta_{bc}$ applies directly:

$$\varepsilon_{mnl}\varepsilon_{nab} = \varepsilon_{nlm}\varepsilon_{nab} = \delta_{la}\delta_{mb} - \delta_{lb}\delta_{ma}.$$

The two Kronecker deltas then collapse the velocity and wavevector indices:

$$\begin{aligned}
\hat{s}(\boldsymbol{k}) &= -k_m \sum_{\boldsymbol{p}} \left[(\delta_{la}\delta_{mb} - \delta_{lb}\delta_{ma})\, p_a \hat{u}_b(\boldsymbol{p})\hat{u}_l(\boldsymbol{k}-\boldsymbol{p})\right] \\
&= \sum_{\boldsymbol{p}} \left[(\boldsymbol{k}\cdot\boldsymbol{p})\, \hat{u}_i(\boldsymbol{p})\hat{u}_i(\boldsymbol{k}-\boldsymbol{p}) - k_i p_j\, \hat{u}_i(\boldsymbol{p})\hat{u}_j(\boldsymbol{k}-\boldsymbol{p})\right] \\
&= -\sum_{\boldsymbol{p}} M_{ij}^{\boldsymbol{k}}(\boldsymbol{p})\, \hat{u}_i(\boldsymbol{p})\hat{u}_j(\boldsymbol{k}-\boldsymbol{p}),
\end{aligned}$$

where $M_{ij}^{\boldsymbol{k}}(\boldsymbol{p}) = k_i p_j - (\boldsymbol{k}\cdot\boldsymbol{p})\delta_{ij}$, exactly as defined in the main text. The geometric content of $\nabla \cdot (\boldsymbol{\omega} \times \boldsymbol{u})$—two derivatives and a cross product—has thus been compressed into the single real tensor $M_{ij}^{\boldsymbol{k}}(\boldsymbol{p})$. Since $S(\boldsymbol{k}) = \langle|\hat{s}|^2\rangle$, the overall minus sign is immaterial and is henceforth absorbed.

#### Step 3. Wick factorization and selection rules

Writing $\hat{s}(\boldsymbol{k}) = \sum_{\boldsymbol{p}} M_{\alpha\beta}^{\boldsymbol{k}}(\boldsymbol{p})\, \hat{u}_\alpha(\boldsymbol{p})\hat{u}_\beta(\boldsymbol{k}-\boldsymbol{p})$ (sign absorbed) and using that $M$ is real,

$$S(\boldsymbol{k}) = \sum_{\boldsymbol{p},\boldsymbol{q}} M_{\alpha\beta}^{\boldsymbol{k}}(\boldsymbol{p}) M_{\gamma\sigma}^{\boldsymbol{k}}(\boldsymbol{q}) \left\langle \hat{u}_\alpha(\boldsymbol{p})\hat{u}_\beta(\boldsymbol{k}-\boldsymbol{p})\hat{u}_\gamma^*(\boldsymbol{q})\hat{u}_\sigma^*(\boldsymbol{k}-\boldsymbol{q})\right\rangle.$$

For the Gaussian absolute equilibrium, Wick's theorem factorizes the fourth-order moment into three pairwise **combinations**:

$$\langle ab\, c^* d^* \rangle = \langle ab \rangle \langle c^* d^* \rangle + \langle ac^* \rangle \langle bd^* \rangle + \langle ad^* \rangle \langle bc^* \rangle.$$

Two ingredients then combine to eliminate two of the three pairings. First, **homogeneity** enforces

$$\langle \hat{u}_i(\boldsymbol{p}) \hat{u}_j^*(\boldsymbol{q}) \rangle = \Phi_{ij}(\boldsymbol{p})\, \delta_{\boldsymbol{p},\boldsymbol{q}}, \qquad \langle \hat{u}_i(\boldsymbol{p}) \hat{u}_j(\boldsymbol{q}) \rangle = \Phi_{ij}(\boldsymbol{p})\, \delta_{\boldsymbol{p},-\boldsymbol{q}},$$

the second identity following from the reality condition $\hat{\boldsymbol{u}}(-\boldsymbol{q}) = \hat{\boldsymbol{u}}^*(\boldsymbol{q})$. Second, the resulting Kronecker deltas act as **selection rules** on the double sum $(\boldsymbol{p}, \boldsymbol{q})$. The three pairings behave as follows.

**Pairing (a) [anomalous; vanishes for $\boldsymbol{k} \neq \boldsymbol{0}$].** This pairing involves the anomalous (un-conjugated) second-order moment $\langle \hat{u}_\alpha(\boldsymbol{p}) \hat{u}_\beta(\boldsymbol{k}-\boldsymbol{p}) \rangle$ and its conjugate partner $\langle \hat{u}_\gamma^*(\boldsymbol{q}) \hat{u}_\sigma^*(\boldsymbol{k}-\boldsymbol{q}) \rangle$, as opposed to the normal correlators $\langle \hat{u}\, \hat{u}^* \rangle = \Phi_{ij}$ that feed pairings (b) and (c). For a real velocity field, the reality condition $\hat{\boldsymbol{u}}(-\boldsymbol{q}) = \hat{\boldsymbol{u}}^*(\boldsymbol{q})$ ties the anomalous correlator to the spectral tensor:

$$\langle \hat{u}_i(\boldsymbol{p}) \hat{u}_j(\boldsymbol{q}) \rangle = \langle \hat{u}_i(\boldsymbol{p}) \hat{u}_j^*(-\boldsymbol{q}) \rangle = \Phi_{ij}(\boldsymbol{p})\, \delta_{\boldsymbol{p},-\boldsymbol{q}},$$

so two un-conjugated modes are correlated only when they sit at opposite wavenumbers, $\boldsymbol{q} = -\boldsymbol{p}$[25]. Setting $\boldsymbol{q} = \boldsymbol{k}-\boldsymbol{p}$, the selection rule $\boldsymbol{p}+\boldsymbol{q} = \boldsymbol{0}$ becomes $\boldsymbol{p}+(\boldsymbol{k}-\boldsymbol{p}) = \boldsymbol{k} = \boldsymbol{0}$: the anomalous factor is non-zero only at the origin $\boldsymbol{k} = \boldsymbol{0}$, and the conjugate partner $\langle \hat{u}_\gamma^*(\boldsymbol{q}) \hat{u}_\sigma^*(\boldsymbol{k}-\boldsymbol{q}) \rangle$ obeys the very same condition.

Since the acoustic wavenumber $k = |\boldsymbol{k}| = f/c > 0$ for any sound of finite frequency, pairing (a) is rigorously absent from the source spectrum (48)—it is not a small correction to be neglected but a term that vanishes identically by the joint action of homogeneity and the reality condition. Physically, this reflects the fact that $|\hat{s}(\boldsymbol{k})|^2$ couples each mode $\hat{\boldsymbol{u}}(\boldsymbol{p})$ only to a conjugate partner $\hat{\boldsymbol{u}}^*(\boldsymbol{k}-\boldsymbol{p})$ (pairings (b) and (c)); an un-conjugated pair $\hat{\boldsymbol{u}}(\boldsymbol{p})\hat{\boldsymbol{u}}(\boldsymbol{k}-\boldsymbol{p})$ would require two co-rotating Fourier amplitudes to be phase-locked at opposite wavenumbers, which homogeneity forbids unless $\boldsymbol{k} = \boldsymbol{0}$[25]. The Gaussian (Wick) structure of the absolute equilibrium[21,30] then guarantees that no residual fourth-order cumulant can revive this term.

**Pairing (b).** $\langle \hat{u}_\alpha(\boldsymbol{p}) \hat{u}_\gamma^*(\boldsymbol{q}) \rangle \langle \hat{u}_\beta(\boldsymbol{k}-\boldsymbol{p}) \hat{u}_\sigma^*(\boldsymbol{k}-\boldsymbol{q}) \rangle$ forces $\boldsymbol{p} = \boldsymbol{q}$ (the second delta, $\boldsymbol{k}-\boldsymbol{p} = \boldsymbol{k}-\boldsymbol{q}$, being then automatic). The $\boldsymbol{q}$-sum collapses, yielding

$$\sum_{\boldsymbol{p}} M^{\boldsymbol{k}}_{\alpha\beta}(\boldsymbol{p}) M^{\boldsymbol{k}}_{\gamma\sigma}(\boldsymbol{p})\, \Phi_{\alpha\gamma}(\boldsymbol{p})\, \Phi_{\beta\sigma}(\boldsymbol{k}-\boldsymbol{p}),$$

i.e. the first bracket of Eq. (48). The index routing is "straight": $\alpha$ connects $M(\boldsymbol{p})$ to $\Phi(\boldsymbol{p})$, $\beta$ connects $M(\boldsymbol{p})$ to $\Phi(\boldsymbol{k}-\boldsymbol{p})$.

**Pairing (c).** $\langle \hat{u}_\alpha(\boldsymbol{p})\hat{u}^*_\sigma(\boldsymbol{k}-\boldsymbol{q})\rangle\langle \hat{u}_\beta(\boldsymbol{k}-\boldsymbol{p})\hat{u}^*_\gamma(\boldsymbol{q})\rangle$ forces $\boldsymbol{p} = \boldsymbol{k}-\boldsymbol{q}$, i.e. $\boldsymbol{q} = \boldsymbol{k}-\boldsymbol{p}$ (the second delta, $\boldsymbol{k}-\boldsymbol{p} = \boldsymbol{q}$, being then automatic). The $\boldsymbol{q}$-sum collapses, yielding

$$\sum_{\boldsymbol{p}} M^{\boldsymbol{k}}_{\alpha\beta}(\boldsymbol{p}) M^{\boldsymbol{k}}_{\gamma\sigma}(\boldsymbol{k}-\boldsymbol{p})\,\Phi_{\alpha\sigma}(\boldsymbol{p})\,\Phi_{\beta\gamma}(\boldsymbol{k}-\boldsymbol{p}),$$

i.e. the second bracket of Eq. (48). Here the index routing is "crossed": $\alpha$ now connects $M(\boldsymbol{p})$ to $\Phi(\boldsymbol{p})$ through $\sigma$, and $\beta$ connects $M(\boldsymbol{k}-\boldsymbol{p})$ to $\Phi(\boldsymbol{k}-\boldsymbol{p})$ through $\gamma$—the $\gamma, \sigma$ indices are swapped relative to pairing (b) because the mode conjugated with $\hat{u}_\alpha(\boldsymbol{p})$ is now $\hat{u}^*_\sigma$ rather than $\hat{u}^*_\gamma$.

Adding pairings (b) and (c) reproduces Eq. (48) exactly. The two surviving terms are thus the two admissible Gaussian combinations of second-order moments; their distinct index routing is precisely what makes $S(\boldsymbol{k})$ sensitive to the off-diagonal (helicity-carrying) structure of $\Phi_{ij}$, and hence to the helicity-induced polarization discussed in the main text.